\documentclass[12pt]{article}

\usepackage{amsmath,amssymb,graphicx} 
\usepackage{epsfig}
\usepackage{pstricks}

\newcommand{\beq}{\begin{eqnarray}}
\newcommand{\eeq}{\end{eqnarray}}
\newcommand{\pI}{\mathsf{p1}}
\newcommand{\pg}{\mathsf{pg}}
\newcommand{\pgg}{\mathsf{pgg}}
\newcommand {\Zb} {\mathbb{Z}} 
\newcommand {\Rb} {\mathbb{R}} 
\newcommand {\id} {\mathbf{1}}
\newcommand{\eq}[1]{eq.~(\ref{#1})}

\newcommand{\centeron}[2]{{\setbox0=\hbox{#1}\setbox1=\hbox{#2}\ifdim
                                        
\wd1>\wd0\kern.5\wd1\kern-.5\wd0\fi
\copy0

\kern-.5\wd0\kern-.5\wd1\copy1\ifdim\wd0>\wd1
                                       \kern.5\wd0\kern-.5\wd1\fi}}
\newcommand{\ltap}{\>\centeron{\raise.35ex\hbox{$<$}}
                               {\lower.65ex\hbox{$\sim$}}\>}
\newcommand{\gtap}{\>\centeron{\raise.35ex\hbox{$>$}}
                               {\lower.65ex\hbox{$\sim$}}\>}

\newcommand\ZZ{\hbox{\zfont Z\kern-.4emZ}}
\font\zfont = cmss10 

\begin{document}
\begin{titlepage}
\begin{flushright}
\end{flushright}

\vskip.5cm
\begin{center}
{\huge \bf 
A Dark Matter candidate from \\ 
\vskip0.2cm
Lorentz Invariance in 6D.
}

\vskip.1cm
\end{center}
\vskip0.2cm

\begin{center}
{\bf
Giacomo Cacciapaglia, Aldo Deandrea {\rm and}
J\'er\'emie Llodra-Perez}
\end{center}
\vskip 8pt

\begin{center}
 {\it Universit\'e de Lyon, F-69622 Lyon, France; Universit\'e Lyon 1, Villeurbanne;\\
CNRS/IN2P3, UMR5822, Institut de Physique Nucl\'eaire de Lyon\\
F-69622 Villeurbanne Cedex, France } \\

{\tt  g.cacciapaglia@ipnl.in2p3.fr,\\ deandrea@ipnl.in2p3.fr, \\jllodra@ipnl.in2p3.fr} 
\end{center}

\vglue 0.3truecm

\begin{abstract}
\vskip 3pt
\noindent
We study the unique 6 dimensional orbifold with chiral fermions where a stable dark matter candidate is present due to 
Lorentz invariance on the orbifold, with no additional discrete symmetries imposed by hand.
We propose a model of Universal Extra Dimensions where a scalar photon of few hundred GeV is a good candidate 
for dark matter.
The spectrum of the model is characteristic of the geometry, and it has clear distinctive features compared to previous 
models of Kaluza-Klein dark matter.
The 5 dimensional limit of this model is the minimal model of natural Kaluza-Klein dark matter.
Notwithstanding the low mass range preferred by cosmology, the model will be a challenge for the LHC due to the 
relatively small splitting between the states in the same KK level.

\end{abstract}

\end{titlepage}

\newpage


\section{Introduction}
\label{sec:intro}
\setcounter{equation}{0}
\setcounter{footnote}{0}

Observations indicate that most of the matter in the universe is dark and of non-baryonic origin with a hint that it could 
be non-relativistic and weakly interacting at late times (cold dark matter). Various potential candidates have been 
suggested, among the most popular, there are weakly interacting massive particles (WIMPs). 
In the following we shall focus uniquely on this possibility. 
This group includes the lightest supersymmetric particle (LSP) obtained by imposing R-parity in supersymmetric 
models~\cite{LSP}; the lightest T-odd particle (LTP) obtained in little Higgs models with T-parity~\cite{LTP}; 
the lightest Kaluza-Klein particle (LKP) which is stable due to a residual parity which may survive in the effective 
theory if special conditions are required for the interactions at the fixed points~\cite{UED5,LKP}.
Apart from phenomenological considerations, which render these symmetries practically compulsory in all these cases, 
the most satisfactory way to introduce a stable particle of this type is to consider a fundamental symmetry. 
From this point of view a residual Kaluza-Klein (KK) parity is the closest candidate, being a remnant of the extra 
dimensional Lorentz invariance after the breaking by compactification. 
However, the compactifications considered so far~\cite{UED5,chiralsquare1} have fixed points or lines which correspond 
to lower dimensional objects (branes) where the extra Lorentz symmetry is absent at all.
Therefore, the KK parity, which is automatic in the bulk of the extra dimensions, has to be imposed by hand on the branes.

We consider these parities unsatisfactory and we want to insist on the idea of keeping a fundamental symmetry as the 
explanation of dark matter: is this possible? 
In the development of extra dimensional model building, the idea of branes and orbifolds with fixed points (both 
suggested by string theory) has proven itself very useful.
For example, one can easily obtain chiral light fermions (from a theory which is inherently non-chiral in 4D).
However, in order to preserve the KK parity, one needs to impose non trivial constraints on the fixed points.
In models based on an interval, one requires the Lagrangian terms localized on the two physically independent end points to 
be identical: obviously, this is not a direct consequence of the compactification!
Moreover, the presence of explicit breaking of Lorentz invariance introduces a large arbitrariness, often neglected, 
in the model, because the localized interaction terms only obey to a smaller subgroup of the extra symmetries.
As an example, divergences which are forbidden in the bulk emerge again via localized counter-terms.
The Ultra-Violet sensitivity of the theory is generically worsened opening a Pandora box of free parameters 
and potentially bad ultraviolet behaviour.
On the other hand, the idea of large compact extra dimensions gives opportunities for model 
building allowing for elegant uses of fundamental symmetries, like in gauge-Higgs unification models~\cite{GHU}, where the freedom to add localized terms is welcome in order to obtain a realistic model.

For those reasons, we will require that Lorentz invariance is only broken globally by the compactification, 
and a KK parity is part of the orbifold, therefore exact and inevitable.
In this spirit, the dark matter candidate is truly a remnant of Lorentz invariance, which we 
shall call by abuse of language, a lightest Lorentz particle (LLP) in the following.
Moreover, all interactions are highly constrained, and the model will be relatively more predictive than standard 
KK parity models. The main hurdle to be passed is the requirement of a chiral spectrum for fermions, crucial to obtain 
the Standard Model in the low energy limit.

Orbifolds are quotient spaces of a manifold modulo a discrete group. 
The one-dimensional orbifolds are the circle $S^1 = \Rb/\Zb$ and the interval $S^1/\Zb_2$: the circle has neither fixed points nor chiral fermions; the interval is the only orbifold with chiral massless fermions, however it possesses two fixed points (the boundaries of the interval).  
If we extend our four dimensional world to include one extra dimension, therefore, there are no compactifications without 
boundaries which allow to obtain chiral fermions in 4D. 
The next step is to consider two extra dimensions. 
In the plane the possible isometries are translations ($t$), reflections, $2\pi/n$ rotations ($r$) with $n=2,3,4,6$ 
and glide-reflections ($g$), which are translations with a simultaneous mirror reflection. 
In 2D there are only 17 fundamental symmetry groups (see for example~\cite{Nilse:2006jv}) which correspond to the 
crystallographic groups in two dimensions, also called the wallpaper groups. The folding of the infinite 
periodic tiling of the plane can be described in a unique way by an orbifold.
 Only 3 of the resulting orbifolds are 
free of boundary or fixed points/lines.
They are the torus $\Rb^2/\pI$, the Klein bottle $\Rb^2/\pg$ and the real projective plane $\Rb^2/\pgg$. 
Our notation refers to the fact that space groups can be defined in a purely algebraic way: instead of specifying a 
representation of the generators, one can list the relations among them, so we have
\begin{eqnarray}
\pI &\simeq& \Zb^2 = \langle t_1 \rangle \times \langle t_2 \rangle\,, \nonumber\\
\pg &=& \langle t_2, g | [g^2,t_2]=0, t_2 g t_2 g^{-1} = \id \rangle \supseteq \Zb^2\,,  \nonumber\\
\pgg &=& \langle r,g | r^2=(g^2r)^2=\id \rangle \supseteq \Zb^2, \Zb_2\,. \nonumber
\end{eqnarray}
Only the projective plane $\Rb^2/\pgg$ allows chiral zero modes for fermions because it contains a $\pi$-rotation, as 
we will discuss in the following. We will therefore consider this geometry as the background for our model. 
One may push this exercise further and consider three or more extra dimensions. 
We will not do it in the following for the simple reason that the number of possible orbifolds increases very fast.
Furthermore, adding more dimensions will lower the effective cutoff of the theory and therefore reduce the validity of 
the effective theory.
The projective plane is the simplest possibility and the most reasonable one for building an effective theory which may 
have a wide validity range in energy.
Here we will consider the minimal case where the Standard Model is embedded in this geometry: the only new parameter, 
neglecting higher order operators, is the size of the extra dimensions which determines the mass scale of the KK resonances.
The geometry also contains two singular points where 4 dimensional terms can be added, and will be required as 
counter-terms for loop divergences: we will consider the most general such Lagrangians, with the spirit that they are 
small corrections (typically one loop level).
This geometry was already proposed in the context of Grand Unification models~\cite{hebecker}, but no explicit 
example has been constructed.

\section{Chiral fermions without fixed points: the real projective plane}
\label{sec:chiral}
\setcounter{equation}{0}
\setcounter{footnote}{0}

The minimal dimension of a fermion $\Psi$ in 6 dimensions is 8 (contrary to the 4 components in 4 and 5 D)~\cite{Polchinski}:
the Clifford algebra contains 6 8$\times$8 Gamma matrices $\Gamma^1 \dots \Gamma^6$.
Moreover, one can define
\beq
\Gamma^7 = \Gamma^1 \Gamma^2 \Gamma^3 \Gamma^4 \Gamma^5 \Gamma^6
\eeq
which allows to define two 6D chiralities via the projectors
\beq
P_\pm = \frac{1}{2} \left( 1 \pm \Gamma^7 \right)\,.
\eeq
Therefore, the minimal spinor representation of the Lorentz group are 4-component chiral fermions $\Psi_\pm = P_\pm \Psi$.
Each one of those 6D-chiral fields contains two 4D Weyl fermions of opposite 4D-chirality.
In order to be more concrete, we will use the following representation for the Gamma matrices~\cite{Polchinski}:
\beq
\Gamma^\mu = \left( \begin{array}{cc} 
\gamma^\mu & 0 \\
0 & \gamma^\mu \end{array} \right)\,, \quad 
\Gamma^5 = \left( \begin{array}{cc} 
0 & i \gamma^5 \\
i \gamma^5 & 0 \end{array} \right)\,, \quad 
\Gamma^6 = \left( \begin{array}{cc} 
0 & \gamma^5 \\
- \gamma^5 & 0 \end{array} \right)\,;
\eeq
and consequently
\beq
\Gamma^7 =  \left( \begin{array}{cc} 
-\gamma^5 & 0 \\
0 & \gamma^5 \end{array} \right)\,, \quad \mbox{and} \quad
P_\pm =  \left( \begin{array}{cc} 
\frac{1}{2} \left( 1\mp \gamma^5 \right) & 0 \\
0 & \frac{1}{2} \left( 1 \pm \gamma^5 \right) \end{array} \right) = 
 \left( \begin{array}{cc} 
P_{L/R} & 0 \\
0 & P_{R/L} \end{array} \right)\,,
\eeq
where $P_L$ and $P_R$ are the projectors on the 4D chiralities.
In this basis, the 6D-chiral fermions can be written as (using the Weyl representation for the 4D Gamma matrices)
\beq
\Psi_+ = \left( \begin{array}{c}
\psi_{L+} \\
\psi_{R+} \end{array} \right)\,, \qquad \Psi_- = \left( \begin{array}{c}
\psi_{R-} \\
\psi_{L-} \end{array} \right)\,,
\eeq
with
\beq
\psi_{L\pm} = \left( \begin{array}{c}
\chi_\pm \\
0 \end{array} \right)\,, \qquad \psi_{R\pm} = 
 \left( \begin{array}{c}
0\\
\bar \eta_\pm 
 \end{array} \right)\,,
 \eeq
where $\psi$ are Dirac spinors,  $\chi$ and $\eta$ are Weyl spinors.

The 6D action for a massless 6D-chiral fermion is
\beq \label{eq:fermaction}
S_{\pm}& =& \int d x_5\, \int dx_6\;  \frac{i}{2} \Big\{ \bar \Psi_\pm \Gamma^\alpha \partial_\alpha \Psi_\pm - \left( \partial_\alpha \bar \Psi_\pm \right) \Gamma^\alpha \Psi_\pm \Big\} = \nonumber\\
& = &  \int d x_5\, \int dx_6\;  \left\{ \phantom{\frac{1}{2}}i \bar \psi_{L\pm} \gamma^\mu \partial_\mu \psi_{L\pm} + i \bar \psi_{R\pm} \gamma^\mu \partial_\mu \psi_{R\pm} + \right. \nonumber \\
& & \left. + \frac{1}{2} \left[ \bar \psi_{L\pm} \gamma_5 (\partial_5 \mp i \partial_6 ) \psi_{R\pm} + \bar \psi_{R\pm} \gamma_5 (\partial_5 \pm i \partial_6 ) \psi_{L\pm} + h.c. \right] \right\}\,;
\eeq
and a mass term
\beq
S_{\rm mass} & = & \int d x_5\, \int dx_6\;  M \Big\{ \bar \Psi_+ \Psi_- + \bar \Psi_- \Psi_+ \Big\} \nonumber \\
 & = & \int d x_5\, \int dx_6\; M \Big\{ \bar \psi_{L+} \psi_{R-} + \bar \psi_{R+} \psi_{L-} + h.c. \Big\}\,.
 \eeq
The only difference between the two 6D-chiralities is a different sign in front of the $x_6$ derivative.
This feature will be important when discussing the parity properties of such fields.

The real projective plane can be defined by a $\pi$-rotation $r$ and a glide $g$, whose action on the coordinates is
\beq
r: \left\{ \begin{array}{l}
x_5 \sim -x_5 \\
x_6 \sim -x_6
\end{array} \right.\,, \qquad
g: \left\{ \begin{array}{l}
x_5 \sim x_5 + \pi R_5\\
x_6 \sim - x_6 + \pi R_6
\end{array} \right.\,.
\eeq
Note that $r^2 = (g^2 r)^2 = 1$; for any field, the allowed parities are therefore
\beq \label{eq:parities}
p_r = \pm 1 \,; \qquad p_g = \pm 1\,.
\eeq
In terms of $r$ and $g$ one can define two translations: 
\beq
t_5 = g^2: \left\{ \begin{array}{l}
x_5 \sim x_5 + 2 \pi R_5 \\
x_6 \sim x_6
\end{array} \right.\,, \qquad
t_6 = (gr)^2: \left\{ \begin{array}{l}
x_5 \sim x_5\\
x_6 \sim x_6 + 2 \pi R_6
\end{array} \right.\,.
\eeq
No Scherk-Schwarz phases can be defined on this geometry: all the fields are periodic.
Note also that we can define another glide
\beq
g' = gr: \left\{ \begin{array}{l}
x_5 \sim - x_5 + \pi R_5\\
x_6 \sim x_6 + \pi R_6
\end{array} \right.\,,
\eeq
under which the fields have parity $p_{g'} = p_g p_r$.
The two radii $R_5$ and $R_6$ are in principle different: here for simplicity we will fix $R_5 = R_6 = R$, and assume 
$R=1$ in all the formulas, except introducing it back when discussing the phenomenology of the model.
The size of the radius will determine the overall mass scale for the KK modes, $m_{KK} = 1/R$.

Let us know discuss the 4D chirality of the fermions.
The glide $g$ changes the sign of one coordinate only, $x_6$: in order to keep the action invariant, 
from eq.~\ref{eq:fermaction}, we see that the two 6D-chiralities are exchanged.
Therefore, the glide requires to start with a non-chiral theory in 6D.
Under the glide, a generic fermion transforms as:
\beq
\Psi (g (x)) = p_g \Gamma_g \Psi (x)\,, \qquad \Gamma_g = \Gamma^6 \Gamma^7\,.
\eeq
The two 6D chiralities are exchanged up to a sign, therefore for both parities a non-chiral 4D massless mode is allowed.
For this reason the Klein bottle, defined by a glide and a translation, does not allow for chiral fermions.

Under the rotation $r$ both extra coordinates change sign, therefore the two 4D chiralities must have opposite parity: 
a zero mode is allowed only for one of the two 4D chiralities.
A generic fermion transforms as:
\beq
\Psi (r (x)) = p_r \Gamma_r \Psi (x)\,, \qquad \Gamma_r =i \Gamma^5 \Gamma^6 \Gamma^7\,;
\eeq
with this definition, $p_r = +1$ corresponds to a left-handed zero mode and $p_r = -1$ to a right-handed one.
Because of eq.~\ref{eq:parities}, for each bulk fermion there is a massless chiral fermion.
The real projective plane is therefore the unique 6D orbifold which allows for chiral fermions without fixed points.
Note also that a bulk mass term is odd under the rotation.
For completeness, the action of the second glide $g'$ on a fermion is defined by:
\beq
\Psi (g' (x)) = p_g p_r \Gamma_{gr} \Psi (x)\,, \qquad \Gamma_{gr} =  i \Gamma^5 \,.
\eeq
Note finally that the presence of a massless mode for each bulk fermion implies that supersymmetry cannot be 
completely broken in this background: in fact, the supersymmetry generator, which is a 6D spinor, obeys the same 
properties, and therefore an unbroken N=1 supersymmetry always survives.

\subsection{KK parities and singularities on the real projective plane}

\begin{figure}[tb]
\begin{center}
\includegraphics[width=11cm]{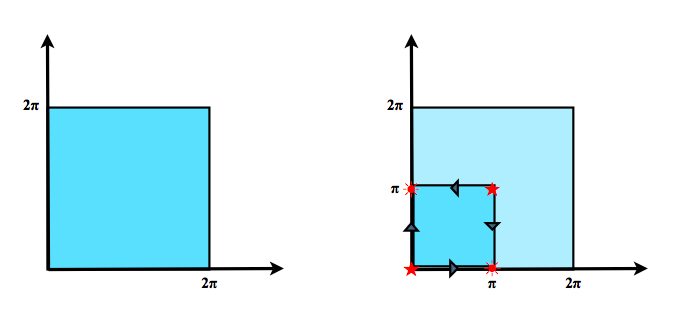}
\end{center}
\caption{\footnotesize Fundamental domain for a torus (left) and real projective plane (right).} \label{fig:orbifold}
\end{figure}

The real projective plane is non-orientable and has no boundaries, however there are two points with conical 
singularities, where localized counter-terms can be added in general. Nevertheless, a KK parity is still preserved 
without any further assumption. Two symmetries can be used to define the orbifold, for example a rotation and a glide.
The rotation has 4 un-equivalent fixed points: $(0,0)$, $(\pi, \pi)$, $(0,\pi)$ and $(\pi,0)$ (see figure~\ref{fig:orbifold}).
The glide, however, transforms $(0,0) \to (\pi, \pi)$ and $(0,\pi) \to (\pi, 0)$ (and viceversa); therefore, no fixed point is present globally, and the the eventual localized interactions on the corners of the fundamental square must be identified in pairs. The identification is not a consequence of an ad-hoc global symmetry acting on the 
UV completion of the model, as it happens in other orbifolds, like the chiral square~\cite{chiralsquare1}, but 
it is part of the orbifold itself: the two identified points are indeed the same point in the 6D space.
A crucial consequence is that a discrete KK parity is left unbroken: it can be identified with a translation 
by $(\pi, \pi)$ (combined with $r$, this is equivalent to a $\pi$-rotation around the center of the square 
$(\pi/2, \pi/2)$, thus it is exactly the same KK parity as in~\cite{chiralsquare1}):
\beq
p_{KK}: \left\{ \begin{array}{l}
x_5 \sim x_5 + \pi \\
x_6 \sim x_6 + \pi
\end{array} \right. \,.
\eeq
Under this transformation, two identified singular points are transformed one into the other, and a generic KK mode 
with momentum $(k,l)$ along the extra directions will acquire a phase $(-1)^{k+l}$: therefore, $(1,0)$ and $(0,1)$ 
identify the lightest tiers of odd particles thus containing a stable Dark Matter candidate, while the $(1,1)$ modes 
can generically decay into zero modes via localized interactions.
Note that the theory also possesses another KK parity
\beq
{p'}_{KK}: \left\{ \begin{array}{l}
x_5 \sim x_5 + \pi \\
x_6 \sim x_6
\end{array} \right.\, ,
\eeq
under which the $(1,1)$ states would be odd and therefore stable.
If ${p'}_{KK}$ is unbroken, the two tiers $(1,0)$ and $(0,1)$ would also pick different parity, and therefore contain 
two independent candidates. However, this symmetry requires that the localized interactions on the two points are 
the same: this is true for terms induced by loops of bulk interactions, however a generic UV completion of the model 
would violate such symmetry.
Not being a fundamental symmetry of the orbifold, we will discard it and assume that it is generally broken.

\section{The Standard Model on the real projective plane}
\label{sec:SM}
\setcounter{equation}{0}
\setcounter{footnote}{0}

Here we will consider a simple extension of the Standard Model on the real projective plane, and study the spectrum of 
the KK modes. More complicated constructions are in principle possible, and we reserve to study them in future publications.
For now we will study the SM gauge group SU(3)$_c \times$ SU(2)$_W \times$ U(1)$_Y$ with a single Higgs scalar doublet, 
and a 6D fermion for each chiral SM fermion (doublets $Q$ and $L$, singlets $U$, $D$, $E$ and possibly the singlet 
neutrino $N$). The lowest order Lagrangian will be the SM one, extended to 6 dimensions.
As we will shortly see, to each SM fields it corresponds a tower of massive resonances organized in tiers of modes, 
labeled by two integers $(l, k)$ which correspond to the discretized momenta along the extra directions, and the field content of each tier will crucially depend on the parities of the fields 
under the orbifold projection. At leading order, all the states in each tier are degenerate with mass determined by the two integers
\beq
m^2_{l,k} = \frac{l^2}{R_5^2} + \frac{k^2}{R_6^2}.
\eeq
Splittings within the modes in each tier can be generated by three mechanisms: the Higgs vacuum expectation value (VEV), 
bulk interaction loop corrections and higher order operators localized on the singular points.
Here we will focus on the simplest case, where, due to the flatness of the metric along the extra coordinates, 
the Higgs VEV is constant: an important consequence is that, due to the orthogonality of wave functions of KK modes 
from different tiers, the Higgs mechanism will not mix the tiers.
Therefore, the KK expansion remains valid and the masses will be shifted, independently on the spin of the field, by 
the SM mass $m_0$, according to the formula: $m^2_{l,k} = l^2 + k^2 + m_0^2$.
On the other hand, the loop corrections do generate level-mixings: being small effects, for the spectrum we will limit 
ourselves to the leading corrections, therefore diagonal terms only.
Note also that the loop induced terms will respect the full global symmetries of the space, and therefore, as an example, 
no splitting and/or mixing between the $(1,0)$ and $(0,1)$ levels will be induced, neither decays of the $(1,1)$ modes 
into SM particles.
Localized terms are generally required as counter-terms for the loop divergences: however, they must respect less symmetries 
than the bulk loops. The only unbroken symmetry will be the KK parity $p_{KK}$.
In the following we will assume that localized terms are small as they are required at one loop level, and we will limit 
ourselves to a leading order expansion in them.

In this section, we will first study the tree level spectrum for generic scalars, gauge fields and fermions on the real 
projective plane, thus identifying the possible field content of each tier, and then include the effect of the Higgs 
mechanism and lowest order localized terms on SM fields.
One loops results for the lightest tier will also be presented because they play a crucial role when discussing the 
Dark Matter relic abundance.

\subsection{Scalars} \label{sec:scalartree}

The action for a scalar field $\Phi$ is (omitting the integral along the un-compact 4 dimensions)
\beq
S_{\mbox{scalar}} = \int_0^{2 \pi} d x_5 \, d x_6\; \Big\{ \partial_\alpha \Phi^\dagger \partial^\alpha \Phi - M^2 \Phi^\dagger \Phi \Big\}\,,
\eeq
$\alpha = 1,\dots 6$; which leads to the equation of motion (EOM)
\beq
\left( \partial_5^2 + \partial_6^2 + p^2 - M^2 \right) \Phi = 0\,,
\eeq
where $p^2 = - \partial_\mu \partial^\mu$.
After Fourier transforming along the two extra coordinates, the field can be expanded in a sum of KK modes, whose wave functions satisfy the above equation with $p^2$ replaced by the mass square of the mode.
The solutions of this equation are usual combinations of sines and cosines (with frequencies determined by the periodicity at $2 \pi$). 
The wave functions can be labelled with the parities under the rotation and glide:
%
\begin{center}
\begin{tabular}{c|cc|l}
   & $p_r$ & $p_g$ & spectrum \\
   \hline
 $  \cos k x_5 \; \cos  l x_6 $ & $+$ & $(-1)^{k+l}$ & $k,l \geq 0$ \\
  $ \sin k x_5 \; \sin l x_6 $& $+$ & $(-1)^{k+l+1}$ & $k,l > 0$\\
$\sin k x_5 \; \cos  l x_6 $& $-$ & $(-1)^{k+l}$ & $k>0\,, l\geq 0$\\
$\cos  k x_5 \; \sin l x_6$ & $-$ & $(-1)^{k+l+1}$ & $k \geq 0\,, l>0$
\end{tabular}
\end{center}
%
The masses are given by the formula
\beq
m_{k,l}^2 = M^2 + k^2 + l^2\,.
\eeq
The mass eigenstates can be labeled by their parity assignment $(p_r,p_g)$ and KK number $(k,l)$.
Here is a full classification of the modes (with normalized wave functions):

\begin{center}
\begin{tabular}{c|c|c|c|c|c|c|}
 $(k,l)$ & $p_{KK}$  & $(++)$ & $(+-)$ & $(-+)$ & $(--)$ \\
 \hline
 $(0,0)$ & $+$ & $\frac{1}{2 \pi}$ & & & \\
 \hline
 $(0, 2l)$ & $+$ & $\frac{1}{\sqrt{2} \pi} \cos 2 l x_6$ & & & $\frac{1}{\sqrt{2} \pi} \sin 2 l x_6$ \\
 \hline
 $(0,2l-1)$ & $-$  & &  $\frac{1}{\sqrt{2} \pi} \cos (2 l-1) x_6$ & $\frac{1}{\sqrt{2} \pi} \sin (2 l-1) x_6$ & \\
 \hline
 $(2k,0)$ & $+$  & $\frac{1}{\sqrt{2} \pi} \cos 2 k x_5$ &  & $\frac{1}{\sqrt{2} \pi} \sin 2 k x_5$ & \\
\hline
 $(2k-1,0)$ & $-$ & & $\frac{1}{\sqrt{2} \pi} \cos (2 k-1) x_5$ & & $\frac{1}{\sqrt{2} \pi} \sin (2 k-1) x_5$ \\
\hline
$(k,l)_{\rm k+l\; even}$ & $+$  & $\frac{1}{\pi} \cos k x_5 \cos l x_6$ & $\frac{1}{\pi} \sin k x_5 \sin l x_6$ & $\frac{1}{\pi} \sin k x_5 \cos l x_6$ & $\frac{1}{\pi} \cos k x_5 \sin l x_6$ \\
\hline
$(k,l)_{\rm k+l\; odd}$ & $-$ & $\frac{1}{\pi} \sin k x_5 \sin l x_6$ & $\frac{1}{\pi} \cos k x_5 \cos l x_6$ & $\frac{1}{\pi} \cos k x_5 \sin l x_6$ & $\frac{1}{\pi} \sin k x_5 \cos l x_6$ \\
\hline
\end{tabular}
\end{center}

\subsection{Gauge fields} \label{sec:gaugetree}

The action for an abelian gauge field is (also valid at quadratic level for non-abelian gauge symmetries)
\beq
S_{\mbox{gauge}} = \int_0^{2 \pi} d x_5\, d x_6\; \Big\{- \frac{1}{4} F_{\alpha \beta} F^{\alpha \beta} - \frac{1}{2 \xi} \left( \partial_\mu A^\mu - \xi (\partial_5 A_5 + \partial_6 A_6) \right)^2 \Big\}\,,
\label{eq:gaugeAction}
\eeq
where $F_{\alpha \beta} = \partial_\alpha A_\beta - \partial_\beta A_\alpha$, and the $\xi$-gauge fixing term is added to eliminate the mixing between $A_\mu$ and the extra polarizations $A_5$ and $A_6$.
The equation of motion for the vector component is
\beq
-\partial^\mu F_{\mu \nu} - \frac{1}{\xi} \partial_\nu \partial^\mu A_\mu + (\partial_5^2 + \partial_6^2 ) A_\nu
=  (p^2 + \partial_5^2 + \partial_6^2 ) A_\nu = 0\,,
\eeq
which is the same as for a scalar field, and we have assumed that each KK mode satisfies the usual 4D equation in $\xi$-gauge:
\beq
-\partial^\mu F_{\mu \nu} - \frac{1}{\xi} \partial_\nu \partial^\mu A_\mu = p^2 A_\mu\,.
\eeq
Once the parities are assigned, the spectrum and wave functions will be the same as for the scalar field (with $M=0$).

The $A_5$--$A_6$ scalar sector is more complicated: in fact, the massive vector modes acquire their longitudinal 
polarization by eating a tower of scalar components provided by a combination of $A_5$ and $A_6$, while another 
combination will give rise to a single tower of physical scalar states.
Moreover, the parities of the scalar components are determined by the fact that they are part of a 6D vector, 
therefore if the vector component $A_\mu$ has parities $(p_r, p_g)$, the parities of $A_5$ and $A_6$ components are 
respectively $(-p_r, p_g)$ and $(-p_r,- p_g)$. In generic $\xi$-gauge, the EOMs are
\beq 
- \partial_\mu^2 A_5 + \xi \partial_5 (\partial_5 A_5 + \partial_6 A_6 ) + \partial_6 (\partial_6 A_5 - \partial_5 A_6 ) &=& 0\,, \label{eq:A5eom}\\
- \partial_\mu^2 A_6 + \xi \partial_6 (\partial_5 A_5 + \partial_6 A_6 ) + \partial_5 (\partial_5 A_6 - \partial_6 A_5 ) &=& 0\,.\label{eq:A6eom}
\eeq
Here we will focus on two gauge choices: the Unitary gauge, where all non-physical degrees of freedom are removed, and 
the Feynman-'t~Hooft gauge, which is more useful for loop calculations.

\subsubsection*{Feynman-'t~Hooft gauge}

In the Feynman-'t~Hooft gauge $\xi = 1$, the equations of motion for $A_5$ and $A_6$ decouple:
\beq
(\partial_5^2 + \partial_6^2 - \partial_\mu^2) A_{5,6} = 0\,,
\eeq
therefore the two components can be treated as two independent scalar fields with proper parities.
The wave functions and masses are the same as in the scalar case presented in detail in Section~\ref{sec:scalartree}.

\subsubsection*{Unitary gauge}

In the unitary gauge $\xi \to \infty$, the combination
\beq
\partial_5 A_5 + \partial_6 A_6 = 0\,. \label{eq:unigauge}
\eeq
The two fields are not independent, and one can therefore expand both fields on the same tower of 4D scalars $A_{(k,l)}$:
\beq
A_5 = \sum \phi_5 (x_5, x_6) A_{(k, l)}\,, \qquad A_6 = \sum \phi_6 (x_5, x_6) A_{(k, l)}\,,
\eeq
with $\partial_5 \phi_5 + \partial_6 \phi_6 = 0$.
Using the latter relation in eq.s~(\ref{eq:A5eom}--\ref{eq:A6eom}), the two wave functions respect the usual EOM of a 
scalar field:
\beq
(p^2 + \partial_5^2 + \partial_6^2 ) \phi_{5/6} = 0\,;
\eeq
spectra and wave functions are again the same as in the scalar case, with the additional constraint from 
eq.~\ref{eq:unigauge}.
In the following tables, we list in detail the masses and normalized wave functions for the 4 possible parity assignments.
In the $(++)$ case the gauge symmetry is unbroken:
\begin{center}
\begin{tabular}{c|c|c|c|c|}
$(k,l)$ & $p_{KK}$ & $A_\mu^{(++)}$ & $A_5^{(-+)}$ & $A_6^{(--)}$ \\
\hline
$(0,0)$ & $+$ & $\frac{1}{2 \pi}$ &  & \\
\hline
$(0, 2l)$ & $+$ & $\frac{1}{\sqrt{2} \pi} \cos 2 l x_6$ & & \\
\hline
$(0, 2l-1)$ & $-$ & & $\frac{1}{\sqrt{2} \pi} \sin (2 l-1) x_6$ & \\
\hline 
$(2k,0)$ & $+$ & $\frac{1}{\sqrt{2} \pi} \cos 2 k x_5$ & & \\
\hline
$(2k-1,0)$ & $-$ & & & $\frac{1}{\sqrt{2} \pi} \sin (2 k-1) x_5$ \\
\hline 
$(k,l)_{\rm k+l\; even}$ & $+$ & $\frac{1}{\pi} \cos k x_5 \cos l x_6$ & $\frac{l}{\pi \sqrt{k^2+l^2}} \sin k x_5 \cos l x_6$ & -$\frac{k}{\pi \sqrt{k^2+l^2}} \cos k x_5 \sin l x_6$\\
\hline
$(k,l)_{\rm k+l\; odd}$ & $-$ & $\frac{1}{\pi} \sin k x_5 \sin l x_6$ & $\frac{l}{\pi \sqrt{k^2+l^2}} \cos k x_5 \sin l x_6$ & -$\frac{k}{\pi \sqrt{k^2+l^2}} \sin k x_5 \cos l x_6$ \\
\hline
\end{tabular}
\end{center}
If the gauge symmetry is broken by the glide(s), case $(+-)$, there is no zero mode in the spectrum:
\begin{center}
\begin{tabular}{c|c|c|c|c|}
$(k,l)$ & $p_{KK}$ & $A_\mu^{(+-)}$ & $A_5^{(--)}$ & $A_6^{(-+)}$ \\
\hline
$(0,0)$ & $+$ &  &  & \\
\hline
$(0, 2l)$& $+$ &  & $\frac{1}{\sqrt{2} \pi} \sin 2 l x_6$  & \\
\hline
$(0, 2l-1)$ & $-$ & $\frac{1}{\sqrt{2} \pi} \cos (2 l-1) x_6$ && \\
\hline 
$(2k,0)$& $+$ & & & $\frac{1}{\sqrt{2} \pi} \sin 2 k x_5$ \\
\hline
$(2k-1,0)$ & $-$ & $\frac{1}{\sqrt{2} \pi} \cos (2 k-1) x_5$ && \\
\hline 
$(k,l)_{\rm k+l\; even}$& $+$ & $\frac{1}{\pi} \sin k x_5 \sin l x_6$ & $\frac{l}{\pi \sqrt{k^2+l^2}} \cos k x_5 \sin l x_6$ & -$\frac{k}{\pi \sqrt{k^2+l^2}} \sin k x_5 \cos l x_6$\\
\hline
$(k,l)_{\rm k+l\; odd}$ & $-$ & $\frac{1}{\pi} \cos k x_5 \cos l x_6$ & $\frac{l}{\pi \sqrt{k^2+l^2}} \sin k x_5 \cos l x_6$ & -$\frac{k}{\pi \sqrt{k^2+l^2}} \cos k x_5 \sin l x_6$ \\
\hline
\end{tabular}
\end{center}
If the gauge symmetry is broken by the rotation (and the glide $g'$), then there is a zero mode living in the $A_5$ component:
\begin{center}
\begin{tabular}{c|c|c|c|c|}
$(k,l)$& $p_{KK}$ & $A_\mu^{(-+)}$ & $A_5^{(++)}$ & $A_6^{(+-)}$ \\
\hline
$(0,0)$& $+$ & & $\frac{1}{2 \pi}$  & \\
\hline
$(0, 2l)$& $+$ & & $\frac{1}{\sqrt{2} \pi} \cos 2 l x_6$ & \\
\hline
$(0, 2l-1)$ & $-$ & $\frac{1}{\sqrt{2} \pi} \sin (2 l-1) x_6$ && \\
\hline 
$(2k,0)$ & $+$&  $\frac{1}{\sqrt{2} \pi} \sin 2 k x_5$ &&\\
\hline
$(2k-1,0)$  & $-$& & & $\frac{1}{\sqrt{2} \pi} \cos (2 k-1) x_5$ \\
\hline 
$(k,l)_{\rm k+l\; even}$& $+$ & $\frac{1}{\pi} \sin k x_5 \cos l x_6$ & $\frac{l}{\pi \sqrt{k^2+l^2}} \cos k x_5 \cos l x_6$ & $\frac{k}{\pi \sqrt{k^2+l^2}} \sin k x_5 \sin l x_6$\\
\hline
$(k,l)_{\rm k+l\; odd}$ & $-$ & $\frac{1}{\pi} \cos k x_5 \sin l x_6$ & $\frac{l}{\pi \sqrt{k^2+l^2}} \sin k x_5 \sin l x_6$ & $\frac{k}{\pi \sqrt{k^2+l^2}} \cos k x_5 \cos l x_6$ \\
\hline
\end{tabular}
\end{center}
Finally, the gauge symmetry can be broken by both rotation and glide $g$; in this case the zero modes resides in $A_6$:
\begin{center}
\begin{tabular}{c|c|c|c|c|}
$(k,l)$& $p_{KK}$ & $A_\mu^{(--)}$ & $A_5^{(+-)}$ & $A_6^{(++)}$ \\
\hline
$(0,0)$& $+$  && & $\frac{1}{2 \pi}$  \\
\hline
$(0, 2l)$& $+$ & $\frac{1}{\sqrt{2} \pi} \sin 2 l x_6$ & &  \\
\hline
$(0, 2l-1)$ & $-$ & & $\frac{1}{\sqrt{2} \pi} \cos (2 l-1) x_6$ & \\
\hline 
$(2k,0)$& $+$ &  & & $\frac{1}{\sqrt{2} \pi} \cos 2 k x_5$ \\
\hline
$(2k-1,0)$ & $-$ & $\frac{1}{\sqrt{2} \pi} \sin (2 k-1) x_5$ & &\\
\hline 
$(k,l)_{\rm k+l\; even}$& $+$ & $\frac{1}{\pi} \cos k x_5 \sin l x_6$ & $\frac{l}{\pi \sqrt{k^2+l^2}} \sin k x_5 \sin l x_6$ & $\frac{k}{\pi \sqrt{k^2+l^2}} \cos k x_5 \cos l x_6$ \\
\hline
$(k,l)_{\rm k+l\; odd}$ & $-$ & $\frac{1}{\pi} \sin k x_5 \cos l x_6$ & $\frac{l}{\pi \sqrt{k^2+l^2}} \cos k x_5 \cos l x_6$ & $\frac{k}{\pi \sqrt{k^2+l^2}} \sin k x_5 \sin l x_6$\\
\hline
\end{tabular}
\end{center}

\subsection{Fermions} \label{sec:fermiontree}

The action for a 6D Dirac fermion in Section~\ref{sec:chiral} leads to the following EOMs for the 4 components:
\beq
i \bar{\sigma}^\mu \partial_\mu \chi_\pm + (\partial_5 \mp i \partial_6) \bar \eta_\pm &=& 0\,, \\
i \sigma^\mu \partial_\mu \bar \eta_\pm - (\partial_5 \pm i \partial_6) \chi_\pm  &=& 0\,;
\eeq
as usual, we expand each component in a tower of 4D Dirac fermions $f_{l,r}$ (the subscript $l,r$ indicate the 4D 
chirality) satisfying the usual Dirac EOMs.
Those first order equations can be decoupled~\cite{Csaki:2003sh}, and each component satisfies the same quadratic 
equation as the scalar field in the previous section.
The solutions are usual combinations of $\sin$ and $\cos$, and the first order EOMs relate the coefficients of the 
two 4D components. The exact form of the solutions will be determined by the parity assignments of the fields.
The rotation gives a different parity to the two 4D-chiralities, and a fermion with $p_r = +$ ($p_r = -$) will have a 
left-handed (right-handed) zero mode.
On the other hand, the glide will relate the two 6D chiralities, so that the four wave functions are not 
independent: the value of the parity under the glide does not play any role on the zero mode spectrum and, as we will 
see, the only requirement is that the SM doublets and singlets have the same glide parity in order to allow Yukawa 
couplings with the bulk Higgs.

For a left-handed fermion, case $(+ \pm)$, the KK modes are given by:
\begin{center}
\begin{tabular}{c|cccc|}
 $(k,l)$& $\chi_+$ & $\chi_-$ & $\bar{\eta}_+$ & $\bar{\eta}_-$ \\
 \hline
 $(0,0)$ & $\frac{1}{2 \sqrt{2} \pi}$ &  $\pm \frac{1}{2 \sqrt{2} \pi}$ & 0 & 0 \\
 $(0,l)$ & $\frac{1}{2 \pi} \cos l x_6$ & $\pm (-1)^l \frac{1}{2 \pi} \cos l x_6$ & $-\frac{i}{2 \pi} \sin l x_6$ & $\pm (-1)^l \frac{i}{2 \pi} \sin l x_6$ \\
 $(k,0)$ & $\frac{1}{2 \pi} \cos k x_5$ & $\pm (-1)^k \frac{1}{2 \pi} \cos k x_5$ & $-\frac{1}{2 \pi} \sin k x_5$ & $\mp (-1)^k \frac{1}{2 \pi} \sin k x_5$ 
 \end{tabular} \end{center} 
while for both $k,l \neq 0$, there are 2 degenerate solutions for each level which can be parameterized as
\beq
\Psi^{(+\pm)} = \left( \begin{array}{c}
\left( a \cos k x_5\, \cos l x_6 + b \sin k x_5\, \sin l x_6 \right) f_l \\
\pm (-1)^{k+l} \left( c \sin k x_5\, \cos l x_6 - d \cos k x_5\, \sin l x_6 \right) \bar{f}_r\\
\pm (-1)^{k+l} \left( a \cos k x_5\, \cos l x_6 - b \sin k x_5\, \sin l x_6 \right) f_l \\
\left( c \sin k x_5\, \cos l x_6 + d \cos k x_5\, \sin l x_6 \right) \bar{f}_r 
\end{array} \right)\,,
\eeq
where we can use the EOMs and normalization condition to fix the coefficients
\beq
\begin{array}{c}
a = \frac{\cos \alpha}{\sqrt{2} \pi} \\
b = \frac{\sin \alpha}{\sqrt{2} \pi} 
\end{array} \qquad \begin{array}{c}
c = - \frac{k \cos \alpha - i l \sin \alpha}{\sqrt{2} \pi \sqrt{k^2 + l^2}} \\
d =  \frac{k \sin \alpha - i l \cos \alpha}{\sqrt{2} \pi\sqrt{k^2 + l^2}} 
\end{array}
\eeq
The two orthogonal states can be obtained by choosing $\alpha = \theta$ and $\alpha = \pi/2 + \theta$, where 
$\theta$ is an arbitrary mixing angle. 
 
For a right-handed fermion, case $(-\pm)$:
\begin{center}
\begin{tabular}{c|cccc|}
 $(k,l)$& $\chi_+$ & $\chi_-$ & $\bar{\eta}_+$ & $\bar{\eta}_-$ \\
 \hline
 $(0,0)$ & 0 & 0 & $\frac{1}{2 \sqrt{2} \pi}$ &  $\pm \frac{1}{2 \sqrt{2} \pi}$ \\
 $(0,l)$ & $-\frac{i}{2 \pi} \sin l x_6$ & $\pm (-1)^l \frac{i}{2 \pi} \sin l x_6$ & $\frac{1}{2 \pi} \cos l x_6$ & $\pm (-1)^l \frac{1}{2 \pi} \cos l x_6$ \\
 $(k,0)$ & $\frac{1}{2 \pi} \sin k x_5$ & $\pm (-1)^k \frac{1}{2 \pi} \sin k x_5$ & $\frac{1}{2 \pi} \cos k x_5$ & $\pm (-1)^k \frac{1}{2 \pi} \cos k x_5$ 
 \end{tabular} \end{center} 
and, for $k,l \neq 0$:
\beq
\Psi^{(-g)} = \left( \begin{array}{c}
\left( a \sin k x_5\, \cos l x_6 + b \cos k x_5\, \sin l x_6 \right) f_l \\
\pm (-1)^{k+l} \left( c \cos k x_5\, \cos l x_6 - d \sin k x_5\, \sin l x_6 \right) \bar{f}_r\\
\pm (-1)^{k+l} \left( a \sin k x_5\, \cos l x_6 - b \cos k x_5\, \sin l x_6 \right) f_l \\
\left( c \cos k x_5\, \cos l x_6 + d \sin k x_5\, \sin l x_6 \right) \bar{f}_r 
\end{array} \right)\,,
\eeq
with
\beq
\begin{array}{c}
c = \frac{\cos\alpha}{\sqrt{2} \pi} \\
d = \frac{\sin\alpha}{\sqrt{2} \pi} 
\end{array} \qquad \begin{array}{c}
a = \frac{k \cos \alpha + i l \sin\alpha}{\sqrt{2} \pi \sqrt{k^2 + l^2}} \\
b =  -\frac{k \sin\alpha + i l \cos\alpha}{\sqrt{2} \pi \sqrt{k^2 + l^2}} 
\end{array} 
\eeq

Note that we have fixed the normalization of the wave functions in such a way that the mass of each $(k,l)$ KK 
level is real, i.e. $m_{(k,l)} = \sqrt{k^2 + l^2}$.

\subsection{6D loop corrections} \label{sec:loops}

The degeneracy of each KK level is removed at loop level: a complete calculation of the shifts is however beyond the scope 
of this paper. In the following we will focus on the modes $(n,0)$ and $(0,n)$ with $n$ odd, in fact this case covers 
the lightest tiers, and the result will be important in the next section to determine the nature of the Dark Matter 
candidate and estimate its relic abundance.
In general, the loop contributions (that we generically label $\Pi$) can be divided in 4 pieces
\beq
\Pi = \Pi_T + p_g \Pi_G + p_g p_r \Pi_{G'} + p_r \Pi_R\,:
\eeq
the first term $\Pi_T$ is the contribution one would get from the same fields on a torus and, after renormalization 
of the bulk kinetic terms, it leaves a finite contribution.
The other three terms correspond to the two glides and rotation, in the sense that their sign depends on 
the parities $(p_r, p_g)$ of the fields running in the loop.
The contribution of the two glides is finite because the glides do not have any fixed points where a counter-term could 
be localized. On the other hand, the rotation does generate divergences which can be cut-off by counter-terms localized 
on the four points left fixed by the rotation, i.e. the two singular points of the orbifold.
The singularities will be equally spread on the two points, because of the extended global symmetries of the bulk 
interactions. In the next section we will discuss the generic structure of the counter-terms: for now we will limit 
ourselves to cutting off the momentum integral in the loop and compute the coefficient on the $\log$ divergent term.
Note also that bulk interactions respect both $p_{KK}$ and $p'_{KK}$, therefore there will be no mixing between the 
states $(n,0)$ and $(0,n)$: in the following we will compute the diagonal corrections, as the off diagonal ones do
generate sub-leading corrections to the spectrum.

\subsubsection{Gauge bosons}

\setlength{\unitlength}{1cm}
\begin{figure}[tb]
\begin{center}
\includegraphics[width=15cm]{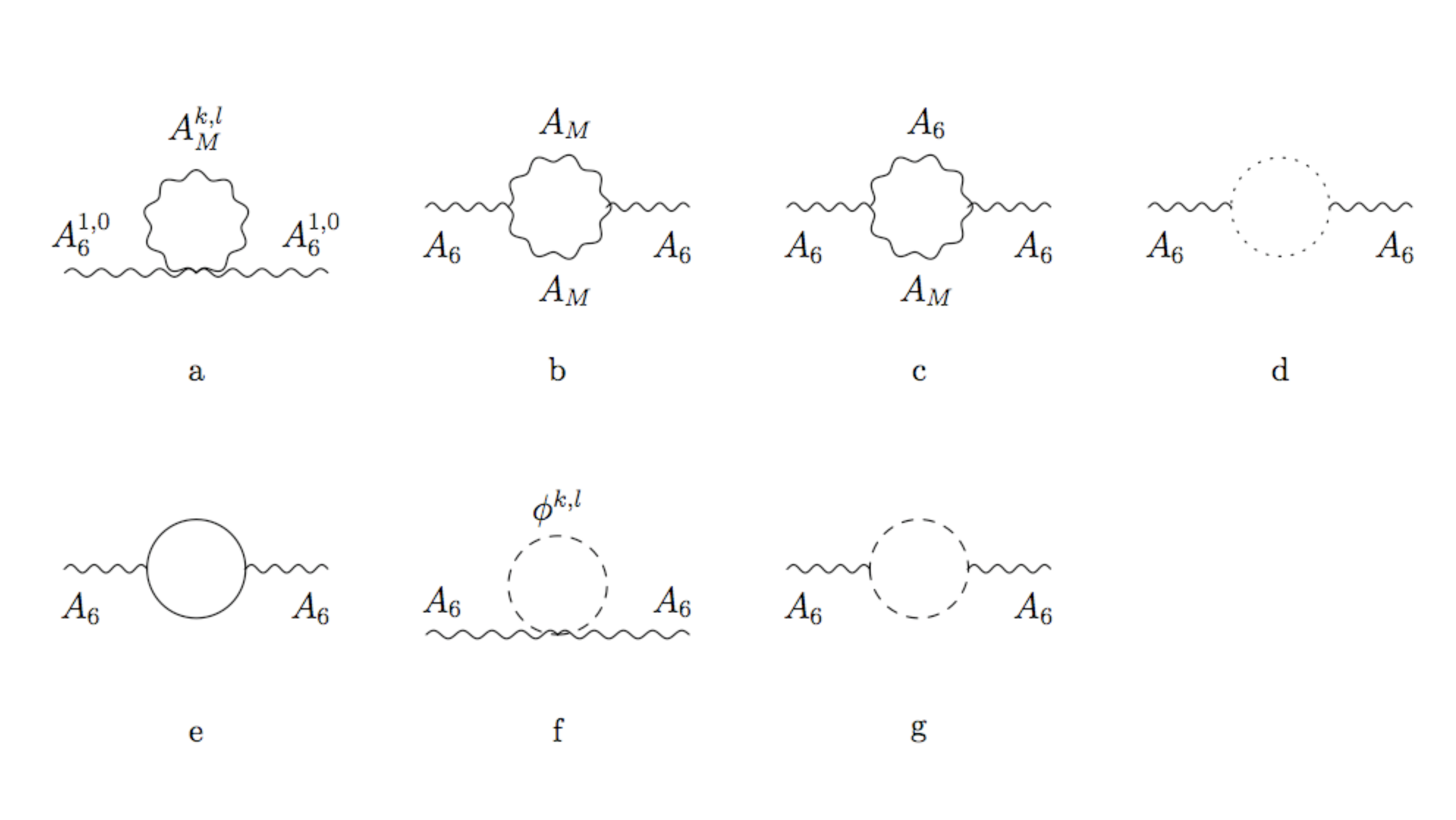}
\end{center}
\caption{\footnotesize One-loop radiative corrections to the gauge scalar self-energy: gauge (a--c), ghost (d), fermion (e) and scalar (f--g) loops.}
\label{fig:loopsGauge}
\end{figure}

The tiers under study contain a gauge-scalar for each SM gauge bosons.
For a generic gauge group, the mass receives corrections from the diagrams listed in Fig.~\ref{fig:loopsGauge}.
We performed the calculation in 3 different ways: we employed a novel method consisting in expanding in KK modes only 
along one direction and using the resummed 5D propagator~\cite{Puchwein:2003jq} in the sum, and we checked the result, 
when possible, using the expansions in winding modes~\cite{winding} and in 6D KK modes~\cite{Cheng:2002iz}.
The methods are summarized in Appendix~\ref{app:loopcorr}, where we explicitly detail the three methods in the 
computation of the scalar loop ``f'' in the figure.
The result of the calculation is summarized in the following table:
\begin{center}
\begin{tabular}{c|cccc|}
$\delta m^2$ gauge scalars   &  & $\times p_g$ & $\times p_g p_r$ & $\times p_r$ \\
 \hline
 a &  $5 T_6$ & $5 \cdot 7 \zeta (3) $ & $3\cdot(7 \zeta (3) + B_1(n))$ &  $3 n^2 \pi^2 L$ \\
 b & $0$ & $0$ & $- 12 B_2(n)$ &  $0$ \\
 c & $- T_6$ & $-3 \cdot 7 \zeta (3)$ & $- (7 \zeta(3) + B_3 (n) )$ &  $5 n^2 \pi^2L$ \\
 d & $0$ & $0$ & $- 2 B_2(n)$ &  $0$ \\
\hline
 e & $-8 T_6$ & $0$ & $0$ &  $0$ \\
\hline
 f & $T_6$ & $7 \zeta (3)$ & $(7 \zeta (3) + B_1(n))$ &  $n^2 \pi^2 L$ \\
 g & $0$ & $0$ & $- 4 B_2(n)$ &  $0$ \\
 \hline
  \end{tabular} \end{center} 
The contributions in the table must be multiplied by a normalization factor $\frac{1}{4} \frac{g^2 C(r)}{16 \pi^4 R^2}$, where $g = \frac{g_6}{2 \pi R}$ is the effective 4D gauge coupling and $C(r)$ is a gauge group factor (defined as $\mbox{Tr} (t_r^a t_r^b) = C(r) \delta^{a,b}$ for a field in the representation $r$ of a non-abelian group running in the loop, and the charge squared for a U(1) ).
In the formula, $T_6$ is the typical sum appearing in the torus compactification~\cite{Cheng:2002iz}
\beq
T_6 = \frac{1}{\pi} \sum_{(k,l) \neq (0,0)} \frac{1}{(k^2 + l^2)^2} \sim 1.92\,,
\eeq
$L = \log \frac{\Lambda^2 + n^2}{n^2}$ is the log divergence associated with the rotation, and the $n$-dependent contributions $B_{1,2,3}$ are small corrections listed in the Appendix~\ref{app:formulae} and coming from heavier modes running in the loop.
Summing over the SM fields, the corrections are:
\beq
\delta m^2_B &=& \frac{{g'}^2}{64 \pi^4 R^2} \left[ - 79 T_6 + 14 \zeta (3) + \pi^2 n^2 L  + B_1 - 4 B_2 \right]\,, \\
\delta m^2_W &=& \frac{{g}^2}{64 \pi^4 R^2} \left[ - 39 T_6 + 70 \zeta (3) + 17 \pi^2 n^2 L + 7 B_1 - 32 B_2 - 2 B_3 \right]\,, \\ 
\delta m^2_G &=& \frac{{g_s}^2}{64 \pi^4 R^2} \left[ - 36 T_6 + 84 \zeta (3) + 24\pi^2 n^2 L + 9 B_1 - 42 B_2 - 3 B_3 \right]\,.
\eeq
Numerically, for n=1, the corrections to the mass $\delta m = \frac{1}{2} \frac{\delta m^2}{m}$ are:
\beq
\delta m_B R &=& (-1.4 + 0.1 L) \cdot 10^{-3} = -0.00094\,,\\
\delta m_W R &=& (-0.4+5.8 L) \cdot 10^{-3} = 0.026\,,\\
\delta m_G R &=& (+0.2+ 28 L) \cdot 10^{-3} = 0.13\,;
\eeq
where we use $\alpha_s (M_Z) = 0.118$, $\alpha (M_Z) = 1/127$, $\sin^2 \theta_W = 0.23$ and $\Lambda = 10$ ($L = 4.6$).

\subsubsection{Fermions}

The corrections to the fermionic Lagrangian for a generic KK mode can be written in general as:
\beq
\delta \mathcal{L} = a_L \bar{\psi} \gamma^\mu p_\mu P_L \psi +  a_R \bar{\psi} \gamma^\mu p_\mu P_R \psi - b \bar{\psi} \psi\,.
\eeq
The wave function renormalizations (in general different for the left-handed and the right-handed components) can be re-absorbed by a field renormalization, so that the shift in the mass (at leading order in the corrections) is:
\beq
\delta m_F = b - m_{n} \frac{a_L + a_R}{2}\,.
\eeq
In the following we will list the contribution of the gauge and scalar (Higgs) loops to the three terms, in units of $\frac{1}{4} \frac{g^2 C_2 (r)}{16 \pi^4 R}$ for the gauge loops (where $C_2 (r) = (N^2-1)/2N$ for a fundamental of SU(N), and the charge squared for a U(1) ) and  $\frac{1}{4} \frac{y^2}{16 \pi^4 R}$ for the Higgs (where $y$ is the effective Yukawa coupling), to be multiplied by the parity of the bosonic field in the loop: 
\begin{center}
\begin{tabular}{c|cccc|}
fermions   &  & $\times p_g$ & $\times p_g p_r$ & $\times p_r$ \\
 \hline
 $n^2\, a_L$ gauge & $0$ & $2 \cdot 7 \zeta (3)$ & $2 \cdot (- 7 \zeta (3) + F_1 (n) + F_2 (n))$ & $0$ \\
 $n^2\, a_R$ gauge & $0$ &$2 \cdot 7 \zeta (3)$ & $- 2 F_2 (n)$ & $0$\\
$n^2\, b$ gauge & $0$ &$4 n \cdot 7 \zeta (3)$ & $-6 n F_2 (n)$ & $4 n^3 \pi^2 L$ \\
\hline
$n^2\, a_L$ scalar & $0$ & $ 7 \zeta (3)$ & $ -7 \zeta (3)+ F_1 (n) + F_2 (n)$ & $n^2 \pi^2 L$ \\
 $n^2\, a_R$ scalar & $0$ &$ 7 \zeta (3)$ & $- F_2 (n)$ & $0$\\
$n^2\, b$ scalar & $0$ & $2 n \cdot 7 \zeta (3)$ & $- n F_2 (n)$ & $n^3 \pi^2 L$ \\
\hline
 \end{tabular} \end{center} 
The $n$-dependent terms are listed in Appendix~\ref{app:formulae}. For a generic fermion in the fundamental representation of SU(2) weak and SU(3) color and with hypercharge $Y_F$:
\begin{multline}
\delta m_F = \frac{1}{64 \pi^2 R n} \left\{ \left( 21 \zeta(3) + 4 n^2 \pi^2 L - F_1 - 6 F_2 \right) \left( Y_F^2 {g'}^2 + \frac{3}{4} g^2 + \frac{4}{3} g_s^2 \right) +  \right. \\ \left.  + \frac{1}{2}  \left( 21 \zeta(3) + n^2 \pi^2 L - F_1 - 2 F_2 \right) y_F^2 \right\}\,. 
\end{multline}
Numerically, at level $n=1$, for each SM fermion we find:
\begin{center}
\begin{tabular}{c|cccccc|}
$\delta m_F R$   & Q & U & D & L & E & N \\
 \hline
 light gen.s & 0.075 & 0.067 & 0.065 & 0.012 & 0.004 & 0 \\
third gen & 0.081 & 0.072  & 0.065 & 0.012 & 0.004 & 0 \\
\hline
 \end{tabular} \end{center} 
where we used the same numerical inputs as for the bosons, and $y_{top} = 1$.

\subsection{Localized operators}

The divergences corresponding to the rotation require the presence of counter-terms localized on the two singular points.
Due to the symmetries of the bulk interactions, the counter-terms should be equal.
However, here we will take a more general approach, and add different terms on the two singularities: in this way we will 
break all additional global symmetries.
In order to leave glide-invariance explicit, we define two localization operators:
\beq
\delta_0 &=& \frac{1}{2} \left( \delta (x_5) \delta (x_6) +  \delta (x_5 - \pi) \delta (x_6 - \pi)  \right)\,, \\
\delta_\pi &=& \frac{1}{2} \left( \delta (x_5) \delta (x_6-\pi) +  \delta (x_5 - \pi) \delta (x_6)  \right)\,,
\eeq
and label the two singular points with a subscript $0$ for the point $(0,0) = (\pi, \pi)$, and $\pi$ for $(0,\pi) = (\pi,0)$.

In general the localized interactions must respect only 4-dimensional Lorentz invariance, and the residual gauge invariance.
For a scalar field like the Higgs many terms can be added including a mass term: here we will neglect this case because 
of the many free parameters and the low phenomenological interest of the Higgs resonances in this model.
For a gauge field the situation is much simpler due to gauge invariance: in order to preserve the residual gauge 
invariance along the extra coordinate, we will write the localized interactions in terms of stress-energy tensor 
components~\cite{Ponton:2005kx}:
\beq
\mathcal{L}_i = \frac{\delta_i}{\Lambda^2} \left( - \frac{r_{1i}}{4} F_{\mu \nu}^2 - \frac{r_{2i}}{2} F_{56}^2 + \frac{r_{5i}}{2} F_{5 \mu}^2 +  \frac{r_{6i}}{2} F_{6 \mu}^2 +  \frac{r_{56i}}{2} F_{5 \mu} F_6^\mu \right)\,;
\eeq
where $i = 0, \pi$, and the cutoff suppression compensates for the dimension of the 6D fields ($2$ for a boson and $5/2$ 
for a spinor). For a standard model gauge boson, with parities $(+,+)$, notice that both $A_{5,6}$ and 
$\partial_{5,6} A_\mu$ vanish on the singular points, therefore $F_{5 \mu} = F_{6 \mu} =  0$ and 
$F_{56} = \partial_5 A_6 - \partial_6 A_5$:
\beq
\mathcal{L}_i = \frac{\delta_i}{\Lambda^2} \left( - \frac{r_{1i}}{4} F_{\mu \nu}^2 - \frac{r_{2i}}{2} ( \partial_5 A_6 - \partial_6 A_5)^2 \right)\,.
\eeq
The first term corrects the kinetic term of the vector bosons, and it also introduces mixing between modes with different 
$(k,l)$. Here we will assume those terms to be small, of the same order as the 1-loop corrections: this is a reasonable 
assumptions because they are in fact counter-terms required by divergences at 1-loop and their coefficient is suppressed 
by the cut-off of the model. Therefore, most off-diagonal terms will give higher order corrections to the masses: this is not the case, however, for tiers that are degenerate like $(k,l)$ and $(l,k)$.
In fact, exchanging the two extra direction is a good symmetry of the real projective plane (this is true for $R_5 = R_6$ 
only, for different radii the degeneracy is removed at tree level).
When expanding the localized terms in KK modes, the $2\times2$ blocks will have equal entries, while the loop contribution 
will be such that the diagonal entries are equal: from this, we see that the block can be diagonalized by the sum and 
difference of the two states~\footnote{The situation is more complicated when $k$ and $l$ are part of a Pitagorean 
triple such that $k^2 + l^2 = n^2$ or quartet with $k^2 + l^2 = n^2 + m^2$: in this cases 3 or 4 states will be 
degenerate. However, this situation only happens for relatively large integers, the smallest ones being 
$(0,5)-(3,4)$, $(5,5)-(1,7)$, $(1,8)-(4,7)$, $(2,9)-(6,7)$, $(0,10)-(6,8)$ and so on. Therefore, the first case happens 
for states of mass $5 \sqrt{2} \sim 7$, which is too close to the cutoff and therefore phenomenologically not interesting: 
for this reason we will not explore the possibility of Pitagorean triples any further.}.
Therefore, we define 
\beq
(k,l)_\pm = \frac{(k,l) \pm (l,k)}{\sqrt{2}}\,, \quad \mbox{with}\; l > k\,, l,k = 0,1, \dots \infty
\eeq
and parameterize the correction to the kinetic term as:
\beq
\mathcal{Z}_{ij} = \delta_{ij} + \frac{z_{ij}}{4 \pi^2 \Lambda^2}\,.
\eeq
In this new basis ($r_{1\pm} = r_{10} \pm r_{1\pi})$:
\begin{center} \begin{tabular}{c|c|c|c|c|c|c|}
$z_{ij}$ & $(0,0)$ & $(0, 2l)_+$ & $(2l,2l)$ & $(2k,2l)_{+}$ & $(2l-1,2l-1)$ & $(2k-1,2l-1)_{+}$ \\
\hline
$(0,0)$ & $ r_{1+}$ & $2 r_{1+}$ & $2 r_{1+}$ & $2 \sqrt{2} r_{1+}$ & $2 r_{1-}$ & $2 \sqrt{2} r_{1-}$ \\
\hline
$(0,2 l')_+$ & $2 r_{1+}$ & $4 r_{1+}$ & $4 r_{1+}$ & $4 \sqrt{2} r_{1+}$& $4 r_{1-}$ & $4 \sqrt{2} r_{1-}$ \\
\hline
$(2l',2l')$ & $2 r_{1+}$ & $4 r_{1+}$ & $4 r_{1+}$ & $4 \sqrt{2} r_{1+}$ & $4 r_{1-}$ & $4 \sqrt{2} r_{1-}$  \\
\hline
$(2k',2l')_{+}$ & $2 \sqrt{2} r_{1+}$ & $4\sqrt{2}  r_{1+}$ & $4\sqrt{2} r_{1+}$ & $8 r_{1+}$& $4\sqrt{2} r_{1-}$ & $8 r_{1-}$ \\
\hline
$(2l'-1,2l'-1)$ & $2 r_{1-}$ & $4 r_{1-}$ & $4 r_{1-}$ & $4 \sqrt{2} r_{1-}$ & $4 r_{1+}$ & $4 \sqrt{2} r_{1+}$  \\
\hline
$(2k'-1,2l'-1)_{+}$ & $2 \sqrt{2} r_{1-}$ & $4\sqrt{2}  r_{1-}$ & $4\sqrt{2} r_{1-}$ & $8 r_{1-}$& $4\sqrt{2} r_{1+}$ & $8 r_{1+}$ \\
\hline
\end{tabular} \end{center}
while $(0,2l)_-$,  $(2k, 2l)_-$, $(2k-1, 2l-1)_-$, $(2k, 2l-1)_\pm$ and $(2k-1,2l)_\pm$ (with $l>k$) are not affected.
The correction to the zero mode will renormalize the gauge coupling:
\beq
g^2 = \frac{g_6^2}{4 \pi^2} \frac{1}{1+\frac{r_{10} + r_{1\pi}}{4 \pi^2 \Lambda^2}} \sim \frac{g_6^2}{4 \pi^2} \left( 1 - \frac{r_{10} + r_{1\pi}}{4 \pi^2 \Lambda^2} + \dots \right)\,;
\eeq 
while the diagonal entries affect the masses:
\beq
m_{(k,l)}^2  = \sqrt{k^2+l^2} \left( 1 - \frac{z_{(k,l)}}{4 \pi^2 \Lambda^2 } + \dots \right)\,.
\eeq
Note also that many states are not affected by the localized terms: this means that the quantum corrections to such 
states are finite at 1-loop.

A similar analysis can be performed for the scalar components of the gauge fields: note that no further mixing between the vector and scalars 
is induced, therefore the tree level bulk gauge fixing term is still appropriate.
In this case, it is the $(k,l)_{\rm ev.}$ modes to be unaffected (in Unitary gauge), and the correction is a mass term
\beq
\delta m^2_{i,j} = m_i m_j \frac{\delta_{ij}}{4 \pi^2 \Lambda^2}\,.
\eeq
In the sum-difference basis, we obtain
\begin{center} \begin{tabular}{c|c|c|c|c|}
$\delta_{ij}$ &  $(0, 2l-1)_+$ & $(0,2l-1)_-$ & $(2k,2l-1)_{+}$ & $(2k,2l-1)_{-}$   \\
\hline
$(0,2 l'-1)_+$ & $4 r_{2\pi}$ & $0$ & $0$  & $4 \sqrt{2}  r_{2\pi}$ \\
\hline
$(0,2 l' -1)_-$ & $0$ & $4  r_{20}$  & $4\sqrt{2} r_{20}$ & $0$ \\
\hline
$(2k',2 l'-1)_{+}$   & $0$ & $4\sqrt{2}  r_{20}$ & $8 r_{20}$ & $0$ \\
\hline
$(2k',2 l'-1)_{-}$  & $4\sqrt{2}  r_{2\pi}$ & $0$ & $0$ & $8 r_{2\pi}$\\
\hline
\end{tabular} \end{center}

This discussion can be easily extended for fermions where, due to the vanishing of one of the two 4D chiral components, 
no mass term can be added and only operators of dimension 6 (like in the gauge boson case) are relevant.

\subsection{Electroweak symmetry breaking: the Higgs VEV}

Another source of mass is the Higgs VEV: here we will assume that the Higgs is $(+,+)$ and its bulk potential contains 
a negative mass, so that a constant VEV is generated for the 6D Higgs scalar field.
Due to the flatness of the VEV along the extra coordinates, it does not induce mixing between different KK tiers, 
therefore the KK expansion presented before can still be utilized to describe mass eigenstates.
At the level $(0,0)$, therefore, we obtain precisely the SM spectrum and the absence of mixing with higher modes also means that the model does not suffer from dangerous tree level corrections to precision electroweak observables like the $S$ and $T$ parameters.
For heavy tiers, the Higgs VEV induces a mixing between the weak neutral bosons, in a similar fashion as in the SM: 
the mixing angle is large because the states are degenerate at tree level, and loop effects will change the value with 
respect to the SM Weinberg angle.
In the fermion sector, only the top will be significantly affected, and the two Dirac fermions corresponding to the left 
and right-handed SM top will be mixed.

\subsubsection{Gauge bosons: general analysis}

The Higgs VEV introduces new mixing between the vectors and some scalar components that must be cancelled by a suitable 
gauge fixing term. For an abelian gauge group, the gauge fixing term in eq.~\ref{eq:gaugeAction} is replaced by
\beq
\mathcal{L}_{\rm \xi-gauge} = - \frac{1}{2 \xi} \left( \partial_\mu A^\mu - \xi (\partial_5 A_5 + \partial_6 A_6 - g v_6 \phi_0) \right)^2\,,
\eeq
when the Higgs is expanded
\beq
H = \frac{1}{\sqrt{2}} \left( v_6 + h + i \phi_0\right)\,.
\eeq
The EOMs of the vector part are modified simply by the addition of a mass term $m_V^2 = g_6^2 v_6^2$.
For the scalars $A_{5,6}$ and $\phi_0$, the new EOMs are:
\beq
(p^2 - m_V^2) A_5 + \partial_6 (\partial_6 A_5 - \partial_5 A_6 ) + m_V \partial_5 \phi_0 + \xi \partial_5 (\partial_5 A_5 + \partial_6 A_6 - m_V \phi_0)  &=& 0\,, \\
(p^2 -m_V^2 ) A_6 + \partial_5 (\partial_5 A_6 - \partial_6 A_5 ) + m_V \partial_6 \phi_0 + \xi \partial_6 (\partial_5 A_5 + \partial_6 A_6 - m_V \phi_0)  &=& 0\,, \\
( p^2 + \partial_5^2 + \partial_6^2) \phi_0 - m_V (\partial_5 A_5 + \partial_6 A_6) +  \xi m_V (\partial_5 A_5 + \partial_6 A_6 - m_V \phi_0) &=& 0\,.
\eeq
In the Feynman-'t~Hooft gauge ($\xi = 1$), the EOMs decouple:
\beq
(p^2 - m_V^2 + \partial_5^2 + \partial_6^2 ) \left[ \begin{array}{c} A_5 \\ A_6 \\ \phi_0 \end{array} \right] = 0\,,
\eeq
 and we have three independent towers with masses $m_{(l,k)}^2 = l^2 + k^2 + m_V^2$, the only difference being the parities 
under rotation and glide.
In the Unitary gauge, the condition
\beq
m_V \phi_0 = \partial_5 A_5 + \partial_6 A_6
\eeq
holds.
Imposing this condition on the EOMs, we obtain the same decoupled equations as in the Feynman-'t~Hooft gauge, however the 
fields are not independent anymore.
Note that one solution of the constraint has the form:
\beq
 \partial_5 A_5 + \partial_6 A_6 = 0 \quad \mbox{and} \quad \phi_0 = 0\,;
 \eeq
those solutions correspond to the physical scalars described in the previous sections, the effect of the Higgs VEV only appears in the 
extra mass contribution $m_V^2$.
The second independent combination of states which satisfies the condition is a new physical scalar, 
mainly consisting of Higgs component: if the gauge symmetry is unbroken by the orbifold
\begin{center}
\begin{tabular}{c|c|c|c|c|}
$(k,l)$ & $p_{KK}$ & $\phi_0^{(++)}$ & $A_5^{(-+)}$ & $A_6^{(--)}$ \\
\hline
$(0,0)$ & $+$ &  &  & \\
\hline
$(0, 2l)$ & $+$ & $\frac{2 l}{\sqrt{2} \pi\, \sqrt{(2l)^2 + m_V^2}} \cos 2 l x_6$ & & $\frac{m_V}{\sqrt{2} \pi\, \sqrt{(2l)^2 + m_V^2}} \sin 2 l x_6$  \\
\hline
$(0, 2l-1)$ & $-$ & &  & \\
\hline 
$(2k,0)$ & $+$ & $\frac{2 k}{\sqrt{2} \pi\, \sqrt{(2k)^2 + m_V^2}} \cos 2 k x_5$ & $\frac{m_V}{\sqrt{2} \pi\, \sqrt{(2k)^2 + m_V^2}} \sin 2 k x_5$ & \\
\hline
$(2k-1,0)$ & $-$ & & & \\
\hline 
$(k,l)_{\rm k+l\; even}$ & $+$ & $\frac{k^2+l^2}{\pi N_{k,l}} \cos k x_5 \cos l x_6$ & $\frac{k m_V}{\pi N_{k,l}} \sin k x_5 \cos l x_6$ & $\frac{l m_V}{\pi N_{k,l}}  \cos k x_5 \sin l x_6$\\
\hline
$(k,l)_{\rm k+l\; odd}$ & $-$ & $\frac{k^2+l^2}{\pi N_{k,l}}  \sin k x_5 \sin l x_6$ & $- \frac{k m_V}{\pi N_{k,l}} \cos k x_5 \sin l x_6$ & $- \frac{l m_V}{\pi N_{k,l}}  \sin k x_5 \cos l x_6$ \\
\hline
\end{tabular}
\end{center}
where $N_{k,l} = \sqrt{(k^2+l^2) (k^2+l^2 + m_V^2)}$.

\subsubsection{Electroweak gauge bosons at the $(1,0)$ and $(0,1)$ tiers}

\begin{figure}[tb]
\begin{center}
\includegraphics[width=7cm]{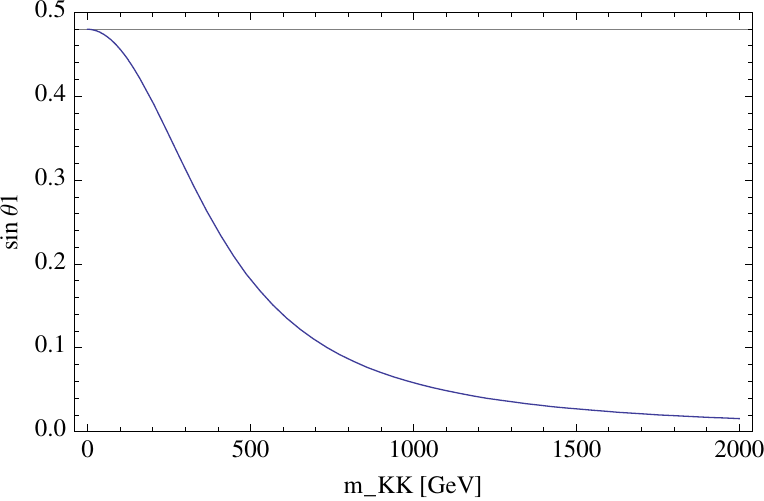} \vspace{0.5cm} \includegraphics[width=7cm]{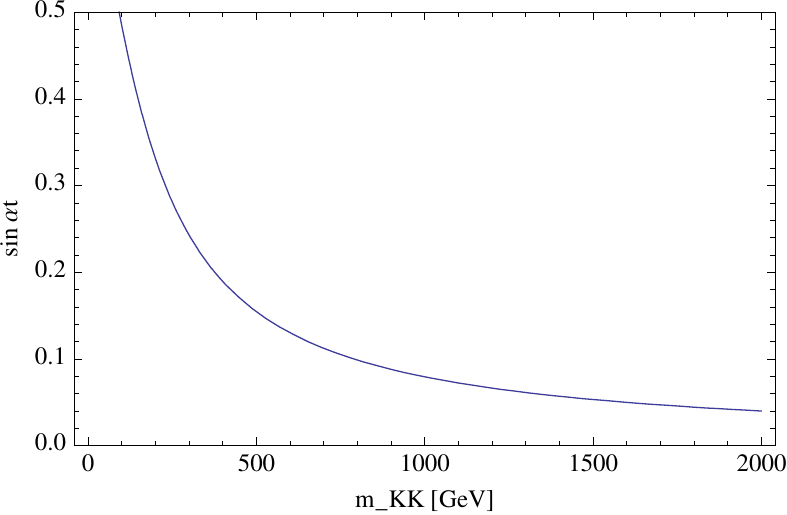}
\end{center}
\caption{\footnotesize Left panel: mixing angle $\sin \theta_1$ between the weak gauge scalars as a function of $m_{KK}$. For zero KK mass we obtain the SM value, 
for large masses the mixing angle vanishes. Right panel: mixing angle in the top sector.} \label{fig:mixing}
\end{figure}

\begin{figure}[tb]
\begin{center}
\includegraphics[width=12cm]{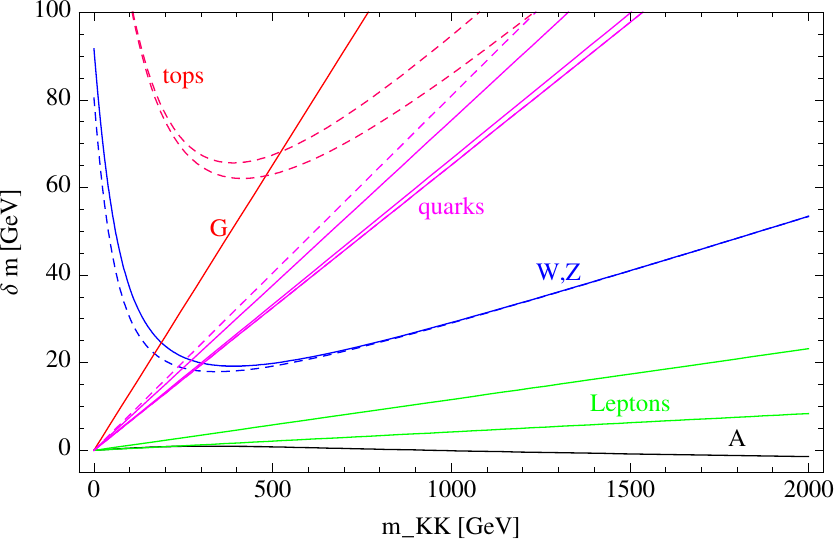}
\end{center}
\caption{\footnotesize Mass splitting between the different states in the tier 1 as a function of $m_{KK}$: in black the 
scalar photon (LLP), in blue the $W$ and $Z$, in solid red the gluon, in green the leptons, in magenta the light quarks, 
in dashed red the tops.} \label{fig:spectra}
\end{figure}

For scalar gauge bosons in the tiers $(n,0)$ and $(0,n)$ with $n$ odd, neglecting the localized operators, the correction 
to the neutral boson mass can be written as:
\beq
\left( \begin{array}{cc} W^3_{n} & B_{n} \end{array} \right) \cdot \left( \begin{array}{cc}
\delta m^2_W + m_W^2 & - \tan \theta_W m_W^2 \\
 - \tan \theta_W m_W^2 &  \delta m^2_B + \tan^2 \theta_W m_W^2
\end{array} \right) \cdot \left( \begin{array}{c} W^3_{n} \\ B_{n} \end{array} \right)\,.
\eeq
Note that neither the Higgs VEV nor the loop corrections mix the two degenerate tiers.
Analogously to the SM, this matrix can be diagonalized by
\beq
\left( \begin{array}{c} Z_{n} \\ A_{n} \end{array} \right) =  \left( \begin{array}{cc}
\cos \theta_n & \sin \theta_n \\
 - \sin \theta_n & \cos \theta_n
\end{array} \right) \cdot \left( \begin{array}{c} W^3_{n} \\ B_{n} \end{array} \right)\,;
\eeq
with mass eigenstates
\begin{multline}
m_{A_n,Z_n}^2 = \frac{n^2}{R^2} + \frac{1}{2} \Big( m_Z^2 + \delta m^2_B + \delta m^2_W  \\
\mp \sqrt{(m_Z^2 + \delta m^2_B - \delta m^2_W)^2 - 4 m_W^2 (\delta m^2_B - \delta m^2_W)} \Big)\,;
\end{multline}
and mixing angle
\beq
\tan \theta_n =  \frac{m_{Z_n}^2  - m_{A_n}^2 + m_Z^2 - 2 m_W^2 + \delta m^2_B - \delta m^2_W}{2 m_W m_Z \sin \theta_W}\,.
\eeq
Note that the mixing angle would be equal to the SM Weinberg angle if $\delta m^2_B = \delta m^2_W$; also, due to the fact 
that the loop corrections grow with the KK mass scale, for large $m_{KK}$ the mixing angle becomes smaller.
A plot of the mixing angle as a function of $m_{KK}$ for the lightest tier is presented in Figure~\ref{fig:mixing}, 
while the splittings are plotted in Figure~\ref{fig:spectra}.
The effect of the localized kinetic terms is easy to include: due to the fact that the bulk contributions (loops and 
Higgs VEV) are the same for the $(n,0)$ and $(0,n)$ tier, they stay diagonal also in the basis $(0,n)_\pm$.
Therefore, the mass eigenvalues and mixing angle are given by the same formulas as before, with 
$\delta m^2_{B,W} \to \delta m^2_{B,W}  + \frac{n^2 r_{2\pi}^{B,W}}{\pi^2 \Lambda^2}$ for $(0,n)_+$ and 
$\delta m^2_{B,W} \to \delta m^2_{B,W}  + \frac{n^2 r_{20}^{B,W}}{\pi^2 \Lambda^2}$ for $(0,n)_-$.

For the charged weak boson, one obtains:
\beq
m_{W^+}^2 = \frac{n^2}{R^2} + \delta m_W^2 + m_W^2\,;
\eeq
the contribution of the localized kinetic terms can be added in the same way as for the neutral bosons.

Similar formulas apply for the vector states in the $(n,0)$ and $(0,n)$ tiers, with $n$ even, and to the $(k,l)$ level: 
however, one needs to take into account the loop-induced mixing between the vector and the scalars, therefore the gauge 
fixing must be redefined at 1-loop.

\subsubsection{Fermions}

The Yukawa couplings are only relevant for he top, therefore we will focus on this case, while the other fermions can be 
obtained by a simple generalization.
The Yukawa couplings can be written as:
\beq
S_{\rm Yukawa} &=&- \int\, d x_5 d x_6\; Y_6\, \bar{\Psi}_{Q} H \Psi_U  + h.c. =\\
  & = & - \int\, d x_5 d x_6\; Y_6\, \left[ \eta_{Q+} H \chi_{U-} + \eta_{Q-} H \chi_{U+} + \bar{\chi}_{Q+} H \bar{\eta}_{U-} + \bar{\chi}_{Q-} H \bar{\eta}_{U+}  \right] + h.c.  \nonumber
\eeq
This term can only be written if $p_r (Q) = - p_r (U)$.
Plugging in the wave functions, and denoting by $q_{l/r}^{(k,l)}$ and $u_{l/r}^{(k,l)}$ the left and right-handed components 
of the $(k,l)$ KK mode, we obtain the corrections to the masses.
For the zero modes:
\beq
\mathcal{L}_{\rm Yukawa (0,0)} = - \frac{p_g (Q) + p_g (U)}{2} \frac{Y_6 v_6}{\sqrt{2}} \bar{q}_l^{(0,0)} u_r^{(0,0)}+ h.c.\,,
\eeq
it is clear that this term is non-vanishing only if the two bulk fermions have the same parity under the glide 
$p_g (Q) = p_g (U) = p_g$, and
\beq
m_{\rm top} = p_g \frac{Y_6 v_6}{\sqrt{2}}\,.
\eeq
For $(l,0)$ and $(0,l)$ modes, we obtain
\beq
\mathcal{L}_{\rm Yukawa (l,0)-(0,l)} = - (-1)^{l} m_{\rm top} \left(\bar{q}_l u_r   -  \bar{q}_r u_l\right)   + h.c.\,.
\eeq
For the $(k,l)$ modes, the situation is more complicated due to the presence of 2 degenerate states: the mass term can 
be written in general as
\begin{multline}
\mathcal{L}_{\rm Yukawa (k,l)} = - (-1)^{k+l} m_{\rm top} (\cos \alpha_Q \cos \alpha_U - \sin \alpha_Q \sin \alpha_U)  \\ \left(\bar{q}_l^{(k,l)} u_r^{(k,l)}   -  \bar{q}_r^{(k,l)} u_l^{(k,l)}\right)   + h.c.\,;
\end{multline}
where the two choices $\alpha_{Q/U} = \theta_{Q/U}, \pi/2 + \theta_{Q/U}$ label the four independent states, and 
$\theta_{Q/U}$ are arbitrary parameters.
If we chose $\theta_Q = - \theta_U$,  two sets of states decouple so that there are no off-diagonal mass entries, 
and the mass matrices reduce to
\beq
\mathcal{L}_{\rm Yukawa (k,l)} = - (-1)^{k+l} m_{\rm top}  \left(\bar{q}_l^{(k,l)} u_r^{(k,l)}   -  \bar{q}_r^{(k,l)} u_l^{(k,l)}\right)   + h.c.\,.
\eeq
This is therefore a general expression valid for all modes.
To find the mass eigenstates, we need to take into account the loop corrections to the $Q$ and $U$ masses.
For the $(1,0)$ and $(0,1)$ tiers:
\beq
\mathcal{L}_{\rm mass} = - \left(\begin{array}{cc} \bar{q}_l & \bar{u}_l \end{array} \right) \cdot   \left(\begin{array}{cc} \frac{1}{R} + \delta m_Q &  - m_{\rm top}\\ m_{\rm top}  & \frac{1}{R} + \delta m_U \end{array} \right) \cdot  \left(\begin{array}{c} q_r \\ u_r \end{array} \right) + h.c.\,.
\eeq
The mass eigenvalues are
\beq
m^2_{t1/2} = \frac{1}{R^2} + m_{\rm top}^2 + \delta m_Q \left(\frac{1}{R} + \frac{1}{2}  \delta m_Q \pm B \right) +  \delta m_U \left(\frac{1}{R} + \frac{1}{2}  \delta m_U \mp B \right)\,,
\eeq
with 
\beq
B = \sqrt{\left( \frac{1}{R} + \frac{\delta m_Q + \delta m_U}{2} \right)^2 + m_{\rm top}^2}\,.
\eeq
The eigenstates are given by
\beq
\left( \begin{array}{c} t_{1l,r} \\ t_{2l,r} \end{array} \right) = \left( \begin{array}{cc} \cos \alpha_t & \pm \sin \alpha_t \\  \mp \sin \alpha_t & \cos \alpha_t \end{array} \right) \left( \begin{array}{c} q_{l,r} \\ u_{l,r} \end{array} \right)\,,
\eeq
and
\beq
\tan \alpha_t = B - \left(\frac{1}{R} + \frac{\delta m_Q + \delta m_U}{2} \right)\,.
\eeq

\section{Dark Matter and collider phenomenology}
\label{sec:pheno}
\setcounter{equation}{0}
\setcounter{footnote}{0}

\begin{figure}[tb]
\begin{center}
\includegraphics[width=12cm]{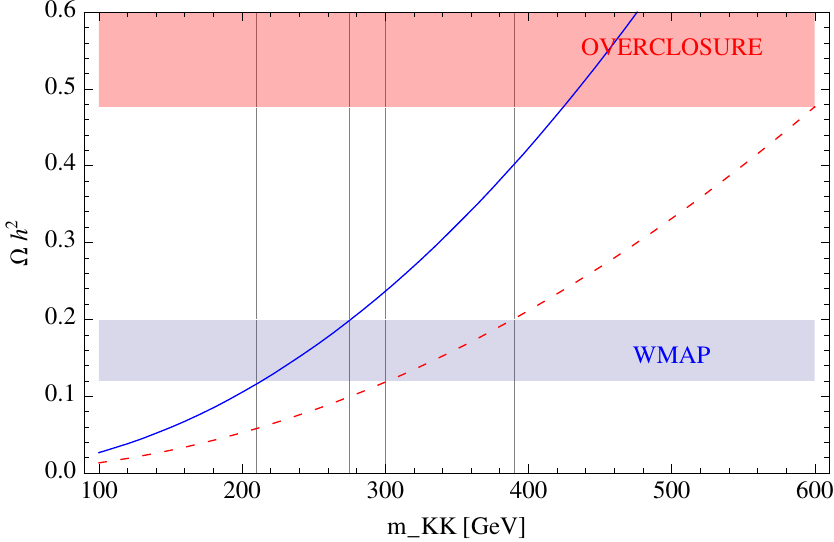}
\end{center}
\caption{\footnotesize Relic abundance of the Dark Matter candidate as a function of $m_{KK}$ for degenerate $(1,0)$ 
and $(0,1)$ levels (solid blue), and for the asymmetric case (dashed red).} \label{fig:relicDM}
\end{figure}

The KK mass, and the Higgs mass, are the main free parameters of the model: calculating the relic Dark Matter abundance 
in this model, one can pin down the cosmologically interesting range for the KK mass.
However, this is nothing but an estimate, because the result is very sensitive to the model of Cosmology and values of 
the cosmological parameters.
In this work, we will assume the standard model of Cosmology and use the approximate formulas in Ref.~\cite{LKP}.
A novel feature with respect to previous works in 6D~\cite{LKP,LKP3} is the smaller splitting between the states in the 
lightest tiers of resonances; therefore, we cannot neglect the co-annihilation with other particles species in the 
tier~\cite{LKP2}.
An average annihilation cross section can therefore be used to estimate the freeze out temperature and the relic density, 
assuming that all the particle species will decay into the LLP after freeze out.
In our calculation, we will neglect electroweak symmetry breaking effects besides the mixing angle in the weak gauge 
boson sector which plays an important role in the calculation due to the relatively small mass splitting between 
the dark matter candidate and the heavier weak gauge resonances.
We also included the main annihilation cross section between all the states in the lightest tier, and assume that the 
localized kinetic terms are negligible (therefore we only included the loop and Higgs contribution to the splittings).
A more detailed study is left for a future publication.
The result for the relic abundance as a function of the KK mass is summarized in figure~\ref{fig:relicDM}: taking 
into account the presence of two degenerate tiers $(1,0)$ and $(0,1)$, we find
\beq
200\;\mbox{GeV} < m_{KK} < 300\;\mbox{GeV}\,;
\eeq
and a limit $m_{KK}<400$ GeV from the over-closure of the Universe.
Note that the two degenerate tiers may be split in the case of asymmetric radii: when one radius is smaller than the other by more than few percent (in particular if the difference in mass is larger that the freeze out temperature, which is typically of order 4\% of the KK mass)
the heavier tier does not contribute significantly to the relic abundance and we obtain a range
\beq
300\;\mbox{GeV} < m_{KK} < 400\;\mbox{GeV}\,.
\eeq
Note also that the mass range can be drastically modified by the localized kinetic terms: for instance, lowering the 
scalar gluon mass will increase the average cross section, and therefore push the preferred mass range to higher values.
However, the presence of such largish terms will also change the splittings of other levels, and modify the branching 
fractions into SM particles: we will not pursue this possibility any further at present.
It is interesting to compare the preferred range with the results in Ref.~\cite{LKP3}, where the chiral square is 
considered. On the chiral square, the splittings are larger than in our case, therefore the lightest state is to a 
good approximation a purely U(1) hypercharge gauge scalar: because of this, the annihilation cross section is smaller, 
and the preferred mass range is lighter (for comparison, for light Higgs $m_H \ll m_{KK}$, they find $m_{KK} < 200$ GeV). 
In our case, the mixing angle is non negligible, and therefore the annihilation cross sections are much larger due to 
the SU(2) interactions.
On the other hand, the co-annihilation with leptons (both singlet and doublet) dilutes the cross section due to the 
large number of leptonic degrees of freedom.
Scalar gluons and quarks do not play a significant role, due to their larger mass.

The inclusion of corrections from the electroweak symmetry breaking, which are important due to the lightness of the 
preferred mass, and resonant annihilation via the Higgs or the $(2,0)$ and $(0,2)$ 
tiers~\footnote{We thank Hiroki Yamashita for pointing this out to us.} can change significantly the range: 
the inclusion of those effects, together with the localized kinetic terms,  in a complete numerical calculation of 
the relic abundance is a subject under study~\cite{arbey}.

Another important issue is the compatibility with electroweak precision measurements~\cite{UED5}: in fact, following 
previous calculations, this low mass range may be excluded once loop corrections to precision observables are taken into 
account. However, a detailed calculation in this specific model has not been performed (and we leave it for a future 
publication); moreover, the low cutoff (naively 10 times the KK mass) means that bulk higher order operators, similar to 
the SM ones, cannot be neglected. Therefore, the prediction power of the model is very limited in this sector: this issue 
is common to all other models of KK Dark Matter; a more detailed analysis is nevertheless required.

In the following we will take the low mass range ($m_{KK} < 400$ GeV) as a ballpark to discuss the phenomenology of the 
model. The main feature of this model compared to previous ones is the relatively small mass splitting among the particles 
in each tier. The states in the first tier will chain decay into the LLP via tree level bulk interactions: the small 
energy available for the SM decay products makes their observation at the LHC challenging.
Below 400 GeV, the heaviest particles are the tops, about 70 GeV heavier than the LLP.
Due to the large mass of the top, the main decay channel will be $t^{(1,0)} \to b W^{(1,0)} \to b W^* A^{(1,0)}$, where 
the virtual $W^*$ converts into a pair of quarks or leptons.
The second heaviest particle is the scalar gluon, and it decays mainly $G^{(1,0)} \to q q^{(1,0)} \to \bar q q A^{(1,0)}$; 
the quarks from light families (from the up to the bottom) decay $q^{(1,0)} \to q A^{(1,0)}$.
The final states from strongly interacting particles, therefore, will always contain 1, 2 or 3 jets and missing transverse 
energy, however the energy of each jet will be rather small, around 20 GeV, making their observation impossible at the 
LHC.
The scalar $W$ and $Z$ will mainly decay to heavy leptons (or quarks for small masses): 
$W^{(1,0)} \to l l^{(1,0)} \to l \nu A^{(1,0)}$, $Z^{(1,0)} \to l l^{(1,0)} \to l^- l^+ A^{(1,0)}$, while leptons will 
decay directly to the LLP $l^{(1,0)} \to l A^{(1,0)}$.
In this case the final state contains leptons, however their typical energy will be small again, less than 20 GeV, 
therefore they will likely escape detection at CMS and Atlas.
In the table we summarize the main decays in the mass range $200 \mbox{GeV} < m_{KK} < 400$ GeV (where MET stands for 
missing transverse energy):

\begin{center}
\begin{tabular}{|l|c|c|c|}
\hline
 & $m_X - m_{LLP}$ & decay mode & final state \\
 & in GeV & & + MET\\
\hline
$t^{(1,0)}$ & $70$ & $b W^{(1,0)}$ & $\begin{array}{c} b j j  \\ b l \nu \end{array}$ \\
$G^{(1,0)}$ & $40$-$70$ & $q q^{(1,0)}$ & $j j$ \\
$q^{(1,0)}$ & $20$-$40$ & $q A^{(1,0)}$ & $j$ \\
$W^{(1,0)}$ & $20$ & $l \nu^{(1,0)}$, $\nu l^{(1,0)}$ & $l \nu$ \\
$Z^{(1,0)}$ & $20$ & $l l^{(1,0)}$ & $l l$ \\
$l^{(1,0)}$ & $<5$ & $l A^{(1,0)}$ & $l$ \\
$A^{(1,0)}$ & $0$ & - & $ $\\
\hline
\end{tabular} \end{center}

The second level accessible at the LHC is the $(1,1)$~\cite{Burdman:2006gy}, with mass $\sqrt{2} m_{KK} = 300 \div 450$ GeV.
Besides chain decays similar to the ones for the lightest tier, the particles in this level can also decay to SM 
particles via localized interactions, only if such interactions break the extra accidental KK parity.
The latter is the only decay mode for the lightest state in such level, therefore, if the level is stable, the same 
phenomenology as the lightest tier applies here, and the observation of such states will be very hard at the LHC.
If the direct decays to SM are possible, one should easily observe resonances without missing energy.
The states in this level can also be singly produced via the same small couplings that induce the decays: as those are 
free parameters in the model, the single-production cross sections and branching ratios cannot be predicted and we will 
not comment on this level any further.

Going up in mass, the next tiers are the $(2,0)$ and $(0,2)$.
In principle, they can decay via bulk interactions in two states from the lightest tier, therefore they would go to 
invisible particles at the LHC.
However, because of the fact that the mass is equal to twice the mass of the lightest tier at three level, the possibility 
that such decays are kinematically open really depends on the loop and Higgs induced splittings, and, in general, those 
decays will be suppressed by the small phase space.
The loops (and kinetic terms) will also induce decays directly into SM particles and single productions, therefore, 
neglecting the localized terms, cross sections and branching fractions for this level can be calculated.
The decays into SM particles will make this level easy to observe in final states without missing energy and with many 
clear resonances.

The first observable missing energy signal will therefore arise at the next level, $(2,1)$ and $(1,2)$, with mass 
$\sqrt{5} m_{KK} = 450 \div 650$ GeV.
They are odd under the KK parity, and due to their mass they can only decay via a loop to a tier-1 particle plus a SM one.
Therefore, the signature for such states will be missing energy plus one SM particle with hundreds of GeV of energy.
Higher modes will repeat this pattern of decays.

Another interesting feature of this model is the possibility of rare but spectacular signals with only one SM particle 
plus missing energy: this is true if the lightest tier is completely unobservable.
For example, one can produce two tier-1 gluons, which radiate a hard gluon: this will generate a mono-jet plus 
missing energy signature~\cite{Alwall:2008va}.
Similarly, one may produce via loop coupling a $(0,1)$ state and a $(2,1)$ state, the latter decaying into $(0,1)$ plus 
a SM state, the only visible particle in the final state. 
More rare but spectacular signals may involve a single charged lepton, however those will be extremely rare due to 
the weak cross section but effectively background free due to the apparent violation of the electric charge.

The model is being implemented in the FeynRules package~\cite{FeynRules} and it will be made publicly available, which will 
allow to interface the Lagrangian of the model to many Monte Carlo tools, and therefore study in detail the phenomenology 
simply sketched in this section.

\subsection{The "5D" limit}

The symmetries of the real projective plane allow to define two different radii along the two directions (this is not 
possible on the chiral square).
Doing so, we would break the symmetry that exchanges the two directions, and remove the degeneracy between levels 
$(n,k)$ and $(k,n)$.
In particular one can take a ``5D'' limit by sending one of the radii, say $R_6$, to zero: in this limit, the model 
collapses to a 5D model where the KK parity is imposed on the boundaries thanks to the extended symmetry of the 6D 
completion. However, the effective model still differs from the 5D case in a crucial way: the limit removes all the modes 
that carry momentum along the $x_6$ direction, so that the structure of the tiers will closely resemble the 5D case, 
however the field content of each level will be very different. This can be rephrased saying that in the 5D limit,  
the 6D physics does not completely decouple: even if the modes that carry momentum along the $x_6$ direction decouple 
from the physical spectrum, the field content and spectrum of the remaining physical particles is in general different 
from the one of a 5D model. On general grounds, this is the consequence of the fact that the starting point geometry is not 
factorized with respect to the extra dimensions, giving rise to a ``geometrical'' non-decoupling of the particle content 
and spectrum of the theory.

Numerically the splittings can be very different from the 6D case. Nevertheless, we can estimate an indicative mass range from the asymmetric case in Figure~\ref{fig:relicDM}, where only one of the lightest tiers is taken into account: the preferred range is $300$-$400$ GeV.
The range is lighter that in the usual UED 5D scenario (around $500\div 600$ GeV~\cite{LKP}) because of the different spin of 
the LLP, and because of the co-annihilation with singlet and doublet leptons.
The phenomenology of the model will be similar to the one of the symmetric 6D one, except for the absence of some 
of the levels (note that the model will appear as a ``5D'' one at the LHC as long as $1/R_6 >$ few TeV).

This example shows that even the minimal 5D model is not unique, and its structure and phenomenology may depend crucially 
on the presence of more dimensions which are too small to be observed.
Our model, in the asymmetric limit, is therefore the minimal 5D model where the KK parity is naturally present, and 
the spin of the LLP can be considered a prediction.

\section{Conclusions}
\label{sec:concl}
\setcounter{equation}{0}
\setcounter{footnote}{0}

We presented a class of models of Dark Matter in 6D where the presence of a Dark Matter candidate does not follow by an ad-hoc 
discrete symmetry, but it is a direct consequence of the topology of the compactification.
The real projective plane is in fact the unique orbifold in 6D that allows both for chiral fermions and the absence of 
fixed points/lines, the latter ensuring the presence of an unbroken parity.
Such parity, a relic of 6D Lorentz invariance, is exact even after including the effect of generic higher order operators 
localized at the two singular points of the compact space.
Even though the model looks very similar to previous proposals in 6D (the chiral square), the topology of the compact space 
crucially affects the loop corrections to the spectrum and the structure of localized operators.
We computed the one loop splitting of the lightest level, and found that the Dark Matter candidate is a scalar photon, 
with a mixing angle smaller that the Weinberg angle.
The main difference with the chiral square is that the splittings in mass within the level tend to be smaller, therefore 
one needs to take into account co-annihilation with leptons and weak gauge bosons and the weak mixing angle cannot be 
neglected.
An estimate of the relic abundance leads to a mass range $200 < m_{KK} < 300$ GeV ($300 < m_{KK} < 400$ GeV in the 
asymmetric case).
In this range, the splittings are such that the observation of the particles in the lightest tier is virtually impossible 
at the LHC as the SM decay products are too soft.
Higher levels are however easy to observe, as they may decay into SM particles without missing energy.
The first missing energy signals may come from the level $(2,1)$ with mass $\sqrt{5} m_{KK}$.
The small splittings also allow for rare but spectacular events with apparent charge non-conservation, like for instance 
mono-jet or mono-lepton plus missing energy, due to the fact that the decay products of the other quark or lepton are too 
soft to be detected.

The model also has a limit where one of the two dimensions is much smaller that the other, so that effectively we have one 
extra dimension with the KK parity imposed by the 6D completion.
Even though only one extra dimension is visible, the phenomenology of the model is very different from the usual 5D case, 
because particle content and mass splitting are very different.
In particular, the KK odd levels do not have vector fields but scalars instead, and the Higgs resonances are missing. 
The DM candidate would be therefore a scalar instead of a vector massive photon.
This is an example of a model in extra dimensions where the presence of dimensions at energies above the TeV scale can 
affect drastically the phenomenology.
Moreover, the 5D model we present is the minimal model of KK Dark Matter in 5D where the KK parity arises naturally, and 
the prediction is the different spin of the lightest stable particle.

\section*{Acknowledgments}
We thank Michele Papucci for useful discussions and comments.
We also thank the Ecole de Physique in Les Houches where the preliminary results of this work were presented during the workshop ``Physics at TeV colliders'', II session, and were we had useful discussions with Renaud Bruneliere, Christophe Grojean, Hiroki Yamashida, Jay Wacker.
The research is supported in part by the ANR project SUSYPHENO (ANR-06-JCJC-0038).

\appendix

\section{Appendix: loop corrections to the masses}
\label{app:loopcorr}
\setcounter{equation}{0}

The calculation of the loop corrections to the masses can be performed using three methods.
First, we can use the expansion of the 6D propagator in winding modes~\cite{winding}: in this way it is straightforward to renormalize the 6D kinetic terms, which corresponds to removing the contribution of the zero winding modes~\cite{Cheng:2002iz}.
However, the calculation is challenging in general due to the presence of Bessel functions in the expansion.
The second way is to expand in KK modes along one direction, and use the resummed 5D propagator~\cite{Puchwein:2003jq} along the other: the advantage is clear when computing corrections to $(n,0)$ modes, where conservation of momentum along the second extra direction simplifies the sum over the KK number in the propagators.
Finally, one can use the usual KK expansion~\cite{Cheng:2002iz}: in this case a more sophisticated technique is required to renormalize the kinetic term.

As an illustration of the three techniques, we will detail the explicit calculation of the scalar loop ``f'' from Figure~\ref{fig:loopsGauge} for the $A_6$ scalar modes $(n,0)$ with $n$ odd.
The results from the other loops has been calculated using at least two of those techniques.

\subsection{6D winding modes method}

A 6D scalar field satisfies the following equation of motion : 
\beq
\left(-\partial_\mu \partial^\mu +\partial_5^2+\partial_6^2\right)\phi=0\,.
\eeq
It is convenient to calculate the propagator in a mixed momentum representation in the un-compactified 4D and position space along the extra directions.
The propagator is therefore the Green function of the following operator
\beq
\left(p^2 +\partial_5^2+\partial_6^2\right) G^{6D}_S (p,x_5-{x'}_5,x_6-{x'}_6) = i \delta(x_5-{x'}_5)\delta(x_6-{x'}_6)\,.
\eeq
The solution~\cite{winding} , defining $p=\sqrt{p^2}$ and $\overrightarrow{y}=(x_5,x_6)$, is:
\beq
G^{6D}_S (p,\overrightarrow{y}-\overrightarrow{y}') = \frac{1}{4}H^{(1)}_0(p\ \arrowvert \overrightarrow{y}-\overrightarrow{y}'\arrowvert) \,,
\eeq
where $H^{(1)}_0$ is the zero order Hankel function of first kind.
The propagator on the real projective plane is given by~\cite{Georgi:2000ks}
\begin{multline} \label{eq:orbprop}
G^{orb}_S (p, \overrightarrow{y},\overrightarrow{y}')=\frac{1}{4} \sum_{\overrightarrow{\Omega}}\; \left[ G^{6D}_S (p,\overrightarrow{y}-\overrightarrow{y}' + \overrightarrow{\Omega}) + p_g\, G^{6D}_S (p, \overrightarrow{y}-g(\overrightarrow{y}')+\overrightarrow{\Omega}) \right. \\
\left.+ p_r \, G^{6D}_S (p, \overrightarrow{y}-r(\overrightarrow{y}')+\overrightarrow{\Omega}) + p_r \ p_g \, G^{6D}_S (p, \overrightarrow{y}-r*g(\overrightarrow{y}')+\overrightarrow{\Omega})\right]
\end{multline}
where $\overrightarrow{\Omega}=(2\pi n_1,2 \pi n_2)$ with $(n1,n2)\in \mathbb{Z}^2$ forces translation invariance ($(n_1, n_2)$ are the winding modes), $p_r$ and $p_g$ are the parities of the scalar field under rotation and glide and $f(\overrightarrow{y}')$ are the transformed of the point $y'$ under the transformation $f$.  

The loop correction to the $A_6$ propagator is given by
\beq
i\Pi^{66} &=& 2ig_6^2 C(r_s)  \eta_{66} \int \frac{d^4 k}{(2\pi)^4} \int d\overrightarrow{y}\;  G^{orb}_S (k,\overrightarrow{y},\overrightarrow{y}) A_6^{(n,0)}(q,\overrightarrow{y}) A_6^{(n,0)}(q,\overrightarrow{y}) \nonumber \\
&=& i N\, (2 \pi)^2 \int d^4 k \int d\overrightarrow{y}\;  G^{orb}_S (k,\overrightarrow{y},\overrightarrow{y}) A_6^{(n,0)}(q,\overrightarrow{y}) A_6^{(n,0)}(q,\overrightarrow{y}) \label{eq:scalartad}
\eeq
where $A_6^{(n,0)} (q,\overrightarrow{y}) = \frac{1}{\sqrt{2} \pi} \sin n x_5$ is the wave function of the external field, $\eta_{66} = -1$ is a metric factor and $g_6^2 = (2 \pi)^2 g^2$ is the 6D gauge coupling.
In the following, in order to simplify the notation, we will always omit the normalization factor  $N = \frac{2 g^2 C(r_s) \eta_{66}}{16 \pi^4}$.
The correction $i\Pi^{66}$ can be split into four terms whose signs depend on the parity of the scalar field under the symmetries of the space:
\beq
\Pi^{66}=\Pi_T + p_g\ \Pi_G + p_r\ \Pi_R+p_g p_r\ \Pi_{G'}\,;
\eeq
$\Pi_T$ is the contribution we would obtain on a torus of same radii and it is finite after the kinetic term renormalization, the other 
three terms are generated by the symmetries of the orbifold and we do expect a $\log$ divergence arising in $\Pi_R$ due to the fixed 
points of the rotation.

\subsubsection*{Torus}

The torus contribution is given by the first term in \eq{eq:orbprop} plugged in \eq{eq:scalartad}:
\beq
\Pi_T& =& \frac{N}{4} 4 \pi^2\ \int d^4k \sum_{\overrightarrow{\Omega}} \frac{1}{4} H^{(1)}_0(k\ \arrowvert \overrightarrow{\Omega}\arrowvert) \nonumber \\
&=&  \frac{N}{4} 4 \pi^3\ \int_0^{\infty} dk_E  \sum_{(n1,n2)\in Z^2} \ \ k_E^3\ K_0(2 \pi k_E \sqrt{n_1^2+n_2^2})
\eeq
where $K_0$ is the K-Bessel function of zero order and we have performed the Wick rotation to write the last integral in Euclidean space.
The zero winding mode $(n_1,n_2)=(0,0)$ contribution is UV divergent, however such divergence is the same we would get in the limit of un-compactified space, therefore it can be absorbed by a wave function renormalization of the 6D field.
Removing the $(0,0)$ mode from the sum and integrating in $k$
\beq
\int_0^{\infty} dk_E  \ k_E^3\ K_0( k_E a) = \frac{4}{a^4}
\eeq
we obtain:
\beq
\Pi_T = \frac{N}{4}\ T_6 \quad \mbox{with} \quad T_6 = \frac{1}{\pi}  \sum_{(n1,n2)\neq (0,0)} \frac{1}{(n_1^2+n_2^2)^2}\sim 1.92\,.
\eeq

\subsubsection*{Glides}

From the second term in the propagator \eq{eq:orbprop}
\beq
\Pi_G = \frac{N}{4} 4 \pi^2  \int d^4k\int d\overrightarrow{y} \sum_{ \overrightarrow{\Omega}} \frac{1}{4} H^{(1)}_0(k\ \arrowvert \overrightarrow{y}-g(\overrightarrow{y}) +\overrightarrow{\Omega}\arrowvert) \frac{\sin^2 n x_5}{2 \pi^2}\,.
\eeq
As the glide does not change sign to the $x_5$ component, the Hankel function does not depend on $x_5$ and its integral will lead to the normalization of wave functions.
After Wick rotating and integrating in $k$ as before we obtain
\beq
\Pi_G=\frac{N}{4\pi^2} \sum_{(n1,n2)\in \mathbb{Z}^2}\int_0^{2 \pi} dx_6 \frac{1}{\left((n_1-1/2)^2 + (x_6/\pi+n_2-1/2)^2\right)^2}\,,
\eeq
where we numerically checked that
\beq
\frac{N}{\pi^2} \sum_{(n1,n2)\in \mathbb{Z}^2}\int_0^{\pi} dx_5 \frac{1}{\left((n_1-1/2)^2 + (x_6/\pi+n_2-1/2)^2\right)^2} = 7 \zeta(3)\,.
\eeq

One obtains a similar expression for the other glide $\Pi_{G'}$, now $x_6$ can be easily integrated out and we are left with
\beq
\Pi_{G'} &=& \frac{N}{4\ \pi^2}  \sum_{(n1,n2)\in \mathbb{Z}^2}\int_0^{2 \pi} dx_5 \frac{1-\cos 2 n x_5}{((x_5/\pi+n_1-1/2)^2 + (n_2-1/2)^2)^2} \\
&=& \frac{N}{4} \left( 7 \zeta(3) + B_1 (n) \right)\,, \nonumber
\eeq
where we again numerically checked that the $n$-dependent term corresponds to the function in Section~\ref{sec:loops}.

\subsubsection*{Rotation}

The last contribution is coming from the rotation part: after the Wick rotation
\beq
\Pi_R=  \frac{ N}{4} 2\pi\ \int_0^{\infty} dk\int d\overrightarrow{y} \sum_{ \overrightarrow{\Omega}} k^3\ K_0(k\ \arrowvert 2\overrightarrow{y} +\overrightarrow{\Omega}\arrowvert) \sin^2 n x_5\,.
\eeq
To extract the divergent part, we cut-off the 4D momentum $k$ at a scale $\Lambda$, and numerically integrated over the compact space and summed.
One can therefore show that the integral is equal to
\beq
\Pi_R=  \frac{N}{4} \quad  n^2 \pi^2 \log{\frac{\Lambda^2+n^2}{n^2}}\,.
\eeq
From the integral form, one can see that the divergences appear when $\arrowvert 2\overrightarrow{y} +\overrightarrow{\Omega}\arrowvert = 0$: 
those points are indeed the fixed points of the rotation, i.e. the corners of the fundamental square.
In his notation, their geometrical origin is clear.

\subsection{6D mixed propagator method}

Using the full 6D propagator is complicated by the fact that one needs to deal with Bessel functions and re-sum a double sum.
On the other hand, 5D propagators can be easily handled, in fact a generic scalar propagator takes the simple form:
\beq
G^{5D}_S (\chi_m ,y-y') = \frac{i \cos \chi_m (\pi-|y-y'|)}{2\ \chi_m\ \sin \chi_m \pi}\quad \mbox{where}\quad \chi_m=\sqrt{k^2-m^2}
\eeq
and $m$ is the 5D mass of the scalar field.
One can therefore exploit this form by expanding in KK modes along one of the extra dimensions, say $x_6$, and write the 6D propagator in terms of resummed 5D propagators:
\beq \label{mixte_propag}
G^{6D}_S (k,\overrightarrow{y}-\overrightarrow{y}') = \sum_{l=-\infty}^\infty  G^{5D}_S (\chi_l, x_5-{x'}_5)\ f^*_l(x_6)f_l({x'}_6)\,,
\eeq
where
\beq 
f_l(x_6) = \frac{1}{\sqrt{2\pi}} e^{ i x_6 l} \quad \mbox{and} \quad \chi_l =\sqrt{k^2-l^2}\,.
\eeq
The $f_l$'s are the wave functions on a circle and $l$ the KK masses for the 5D modes.
This method is extremely powerful, especially to calculate corrections for modes like the $(n,0)$: the fact that the external fields do not carry any momentum along $x_6$, together with the orthonormality of the wave functions $f_l$, allows to easily replace the integral in the coordinate $x_6$ with a sum.
The orbifold propagator and the scalar loop we are considering here are given by eqs.~(\ref{eq:orbprop}) and (\ref{eq:scalartad}).
Contrary to the winding mode method, this one can be extended in a straightforward way to all the other loop diagrams.

\subsection*{Torus}

After integrating over $x_5$ and $x_6$, the torus contribution can be written as:
\beq
\Pi_T = \frac{N}{4} \int d^4k\ \frac{1}{2\pi } \sum_{l=-\infty}^\infty \frac{i \cot\chi_l\pi }{2\ \chi_l}\,.
\eeq
To remove the UV divergence, we decided to renormalized it by regularizing each 5D KK-propagators similarly to \cite{Puchwein:2003jq}, so that:
\beq
\Pi_T = \frac{N}{4} \int d^4k\ \frac{1}{2\pi }  \sum_{l=-\infty}^\infty \frac{i \cot\chi_l\pi -1 }{2\ \chi_l}\,.
\eeq
After Wick rotation, singling out the $l=0$ contribution, one gets:
\beq
\Pi_T &=& \frac{N}{4} \ \left( \zeta (3) +  4\pi^3   \sum_{l=1}^\infty    \int_0^{\infty}  dk_E\ \frac{\ k_E^3 \left(\coth \left[ \pi \sqrt{k_E^2+l^2}\ \right] -1\right)} { 2\ \sqrt{k_E^2+l^2}} \right) \nonumber\\
&=& \frac{N}{4}\ \Delta'\quad \mbox{with}\ \ \Delta' \simeq 1.22 \,;
\eeq
because of the improper regularization scheme, the finite part is different from the previous correct result.
However, the structure is the same, thus providing a powerful way to check the results obtained with the other methods.

\subsection*{Glides}

The contribution of the glides is finite, therefore we do not have the same issue with the renormalization which arose for the torus one.
Under the glide, $x_6$ changes sign and therefore:
\beq
f_l (g(x_6)) = f_l (-x_6 + \pi) = (-1)^l f_{-l} (x_6)\,;
\eeq
due to the orthonormality, the contribution of the $l\neq 0$ modes vanishes, and we are left with
\beq
\Pi_G = \frac{N}{4}\int d^4k \int_{0}^{2\pi } dx_5  \  \frac{i \cos\chi_0(\pi -|x_5-g(x_5)|)}{2\ \chi_0\ \sin \chi_0 \pi } \frac{\sin^2(n x_5)}{2\pi^2}\,.
\eeq
After wick rotation, and integrating in $x_5$, we obtain:
\beq
\Pi_G = \frac{N}{4} \ 2\pi^3 \int_0^{\infty} dk_E  \  \frac{k_E^3 }{k_E\ \sinh \pi k_E } = \frac{N}{4}\  7\zeta(3)\,,
\eeq
that agrees with the result obtained using the winding modes.

Under the second glide
\beq
f_l (gr(x_6)) = f_l (x_6 + \pi) = (-1)^l f_{l} (x_6)\,;
\eeq
therefore all modes contribute, and a $(-1)^l$ factor appears in the sum:
\beq
\Pi_{G'} &=& \frac{N}{4}\int d^4k \int_{0}^{2\pi } dx_5   \sum_{l=-\infty}^\infty \ (-1)^l \frac{i \cos\chi_l(\pi  -|x_5-gr(x_5)|)}{2\ \chi_l\ \sin \chi_l \pi } \frac{\sin^2(n x_5)}{2\pi^2} \nonumber \\
&=& \frac{N}{4}\int_0^{\infty} dk_E\ 2 \pi^2 k_E^3   \sum_{l=-\infty}^\infty \ (-1)^l \frac{  2(k_E^2+l^2)+n^2 }{(k_E^2+l^2) \ (k_E^2+l^2+n^2)}\nonumber\\
&=&\frac{N}{4}\int_0^{\infty} dk_E\ \pi^3\left[  \frac{ 2 k_E^2 }{\sinh \pi k_E }+ \frac{2 k_E^3}{\sqrt {n^2+k_E^2} \sinh \left(\pi\sqrt {n^2+k_E^2} \right)}\right] \nonumber\\
&=&\frac{N}{4}\ \left[  7\zeta(3) +B_1(n)\right]\,.
\eeq
where
\beq
B_1 (n)=2\pi^3  \int_0^{\infty} dp\ \frac{p^3}{\sqrt {n^2+p^2} \sinh \left(\pi\sqrt {n^2+p^2} \right)}
\eeq
From this expression

\subsection*{Rotation}

Similarly to the glide, for the rotation, only the zero mode contributes:
\beq
\Pi_R&=&\frac{N}{4}\int d^4k \int_{0}^{2\pi} dx_5  \ \frac{i \cos\chi_0(\pi  -|x_5-r(x_5)|)}{2\ \chi_0\ \sin \chi_0 \pi } \frac{\sin^2(n x_5)}{2\pi^2} \nonumber \\
&=&\frac{N}{4}\int d^4k \ \frac{i n^2}{k^2\ (n^2-k^2)}\,.
\eeq
After Wick rotation, we regularize the integral by cut off as before:
\beq
\Pi_R=\frac{N}{4}\int_0^{\Lambda} dk_E \ 2\pi^2 k_E \frac{n^2}{(n^2+k_E^2)}
=\frac{N}{4} \quad n^2 \pi^2 \log{\frac{\Lambda^2+n^2}{n^2}}\,.
\eeq

\subsection{6D Kaluza Klein expansions method}

This method, the most commonly used one, makes use of the expansion in 4D KK modes, therefore one needs to compute loops with usual 4D propagators and then sum over the KK momenta of the towers.
However, computing all the necessary couplings between modes can be tedious, and a Fourier transform that goes back to winding modes is necessary for the renormalization of the torus contribution.
Nevertheless, this method can be easily applied to any loop structure.

Here we will again stick to one concrete example.
The contribution of a scalar field with parities $(p_r, p_g)$ can be written as
\beq
\Pi_{p_g\ p_r} = \Pi_T + p_g \Pi_G + p_g p_r \Pi_{G'} + p_r \Pi_R\,;
\eeq
therefore if we calculated the contribution of all 4 parity possibility, we would be able to extract each term:
\beq
\Pi_T&=&\frac{1}{4}\left(\Pi_{++}+\Pi_{+-}+\Pi_{-+}+\Pi_{--}\right)\\
\Pi_G&=&\frac{1}{4}\left(\Pi_{++}-\Pi_{+-}+\Pi_{-+}-\Pi_{--}\right)\\
\Pi_{G'}&=&\frac{1}{4}\left(\Pi_{++}-\Pi_{+-}-\Pi_{-+}+\Pi_{--}\right)\\
\Pi_R&=&\frac{1}{4}\left(\Pi_{++}+\Pi_{+-}-\Pi_{-+}-\Pi_{--}\right)
\eeq
The couplings that enter the loop, $A^6_{(n,0)} A^6_{(n,0)} \phi^\dagger_{(m,l)} \phi_{(m,l)}$ are proportional to $i g^2 \eta_{66}$ with a coefficient that depends on the wave function integrals. We  listed such coefficients in the following table (here $m,l \neq 0$ are intended):
\begin{center} \begin{tabular}{|l|c|c|c|c|}
\hline
  $(p_r\ p_g)$ & $(++)$ & $(+-)$  & $(-+)$  & $(--)$  \\
\hline
$(0,0)$ & $2$ & - & - & - \\
\hline
$(m,0)\ \mbox{m even}$ & $2$ & - & $2$ & - \\
\hline
$(m,0)\ \begin{array}{c} m\neq n \\ \mbox{m odd} \end{array} $ & - & $2$ & - & $2$ \\
\hline
$(n,0)$ & - & $1$ & - & $3$ \\
\hline
$(0,l)$  l even & $2$ & - & - & $2$ \\
\hline
$(0,l)$  l odd & - & $2$ & $2$ & - \\
\hline
$(m,l)\ m\neq n$ & $ 2$ & $2$ & $2$ &  $2$ \\
\hline
$(n,l)$  l even & $ 3$ & $1$ & $1$ &  $3$ \\
\hline
$(n,l)$  l odd &  $1$ & $3 $ & $3 $ &  $1$ \\
\hline
\end{tabular} \end{center}
For example, using the previous table, the correction coming from $\phi_{++}$ and $\phi_{+-}$ running into the loop are:
\begin{align}
i\Pi_{++}=N \, \Big( & \sum_{(m,l)\geq1} 2G(m,l)+\sum_{l\geq1}((-1)^l G(n,l)+2G(0,2l))\nonumber\\
&+\sum_{m\geq1,m\neq n}2G(2m,0)+2G(0,0) \Big)\,,\\
i\Pi_{+-}=N \, \Big(  & \sum_{(m,l)\geq1} 2G(m,l)+\sum_{l\geq1}(-(-1)^l G(n,l)+2G(0,2l-1))\nonumber\\
&+\sum_{m\geq1,m\neq n}2G(2m-1,0)+G(n,0) \Big)\,;
\end{align}
where
\beq
G(m,l)=\int_0^{\infty} dk_E\ \frac{k_E^3}{k_E^2+m^2+l^2}
\eeq
is the integral appearing in the 4D loop.

\subsection*{Torus}

For the torus contribution, we reconstruct a sum over all the KK modes on a torus compactification: following the usual Fourier expansion in the double sum and removing the zero winding mode contribution
\beq
\Pi_T = N \frac{1}{2}  \sum_{(m,l)\in \mathbb{Z}^2} G(m,l)=\frac{N}{4} T_6\,.
\eeq

\subsection*{Glides}

Following the same procedure, the glide contribution can be written as
\beq
\Pi_G=\frac{N}{4} \;  2 \sum_{m\in \mathbb{Z}}(-1)^m G(m,0)=\frac{N}{4} \;  7 \zeta(3)\,;
\eeq
while for the second glide
\beq
\Pi_{G'}= \frac{N}{4} \;  2 \sum_{l\in \mathbb{Z}}(-1)^l( G(n,l)+ G(0,l)) = \frac{N}{4} \;  \left(7 \zeta(3)+B_1(n)\right)\,.
\eeq

\subsection*{Rotation}

For the rotation contribution, the loop calculation gives:
\begin{multline}
\Pi_R=\frac{N}{4} \; 2\left(G(0,0)-G(n,0)\right) = \frac{N}{4} \;  \int_0^\Lambda dk_E\ \left( k_E-\frac{k_E^3}{k_E^2+n^2}\right)\\
=\frac{N}{4} \;  n^2 \pi^2 \log{\frac{\Lambda^2+n^2}{n^2}}\,.
\end{multline}

\section{Appendix: loop integrals}
\label{app:formulae}
\setcounter{equation}{0}

The functions of $n$ appearing in the loop corrections can be expressed in terms of the three following integrals:
\beq
\Phi_1 (n) &=& 2 \pi^3 \int_0^\infty dk\; \frac{k^3}{\sqrt{k^2+n^2} \sinh \pi \sqrt{k^2+n^2}}\,, \\
\Phi_2 (n) &=& 2 \pi^3 \int_0^\infty dk\; \frac{k n (\sqrt{k^2+n^2}-n)}{\sqrt{k^2+n^2} \sinh \pi \sqrt{k^2+n^2}}\,, \\
\Phi_3 (n) &=& 2 \pi^3 \int_0^\infty dk\; \frac{k^3 (\sqrt{k^2+n^2}-n)}{n \sqrt{k^2+n^2} \sinh \pi \sqrt{k^2+n^2}}\,.
\eeq
Those integrals can be computed analytically, and we found
\beq
\Phi_1 (n) &=& 2 \pi n\,\phi_2 (n) - \phi_3 (n)\,,\\
\Phi_2 (n) &=& n (\pi^3 + \pi\, \phi_2 (n) )\,, \\
\Phi_3 (n) &=& \frac{1-2 n^2}{2 n} \pi^3 - 2 \, \phi_3 (n) + \frac{3}{2 \pi n}\, \phi_4 (n)\,;
\eeq
with
\beq
\phi_s (n) = \mbox{Li}_s (e^{2 \pi n}) - 2^s\,  \mbox{Li}_s (e^{\pi n}) - i \frac{2^{s-1} \pi^s}{\Gamma (s)} n^{s-1}\,,
\eeq
where $\mbox{Li}_s$ is the Polylogarithmic function of order $s$ and the imaginary term cancels the imaginary parts of the Polylogs to make a real function of $n$. 
Numerically the integrals are suppressed for large $n$:
\begin{center} \begin{tabular}{|l|c|c|c|}
\hline
   & $n=1$ & $n=2$  & $n=3$ \\
\hline
$\Phi_1$ & $1.43$ & $0.109$ & $0.0067$ \\
\hline
$\Phi_2$ & $0.54$ &  $0.047$ & $0.0030$\\
\hline
$\Phi_3 $ & $1.02$ & $0.037$ &  $0.0015$ \\
\hline
\end{tabular} \end{center}
The functions appearing in the loop corrections are (where we report the numerical value for $n=1$):
\beq
B_1 &=& \Phi_1 \sim 1.43\,, \\
B_2 &=& \Phi_1 - \Phi_2 \sim  0.89\,, \\
B_3 &=& \Phi_1 + \frac{9}{2} \Phi_2 - 3 \Phi_3 \sim 1.83\,, \\
F_1 &=& \Phi_1 - 2 \Phi_2 \sim 0.35\,, \\
F_2 &=& \Phi_2 \sim 0.54\,.
\eeq

\setcounter{equation}{0}


\end{document}